\documentclass[12pt,preprint]{aastex}

\shorttitle{Magnetic turbulence production at SNR shocks}
\shortauthors{Niemiec et al.}
\usepackage{natbib}
\newcommand{\mkp}{}
\newcommand{\mkpa}{}
\newcommand{\tbn}{\tablenotemark}
\begin{document}

\title{Production of Magnetic Turbulence by Cosmic Rays Drifting Upstream 
of Supernova Remnant Shocks}

\author{Jacek Niemiec}
\affil{Instytut Fizyki J\c{a}drowej PAN, ul. Radzikowskiego 152,
 31-342 Krak\'{o}w, Poland}
\email{Jacek.Niemiec@ifj.edu.pl}
\author{Martin Pohl, Thomas Stroman}
\affil{Department of Physics and Astronomy, Iowa State University, Ames,
IA 50011}
\and
\author{Ken-Ichi Nishikawa}
\affil{National Space Science and Technology Center, Huntsville, AL 35805}

\begin{abstract}
We present results of two- and three-dimensional Particle-In-Cell simulations of 
magnetic-turbulence production by isotropic cosmic-ray ions drifting 
upstream of supernova remnant shocks. The studies aim at testing {recent
predictions} of a strong amplification of short-wavelength {\mkp magnetic field}
and at studying the subsequent evolution of the 
magnetic turbulence and its backreaction on cosmic ray trajectories. 
We {\mkp observe} that an oblique filamentary mode grows more rapidly than the non-resonant parallel modes
found in analytical theory, and the growth rate of the field perturbations is much 
slower than is estimated for the parallel plane-wave mode, {\mkp possibly because in our simulations
we cannot maintain $\omega \ll \Omega_i$, the ion gyrofrequency, to the degree required for the 
plane-wave mode to emerge. The {\mkpa evolved}
oblique filamentary mode was also observed in MHD simulations to dominate in
the non-linear phase, {\mkpa when the structures are already isotropic.} We thus}
confirm the generation of the turbulent magnetic
field due to the drift of cosmic-ray ions in the upstream plasma, but 
{\mkp as our main result find that}
the amplitude of the turbulence saturates at about $\delta B/B\sim 1$. 
The backreaction of the magnetic turbulence on the particles leads to an 
alignment of the bulk-flow velocities of the cosmic rays and the background medium. 
{This} is an essential characteristic of
cosmic-ray modified shocks: the upstream flow speed is continuously changed 
by the cosmic rays. {The deceleration of the cosmic-ray drift and the simultaneous
bulk acceleration of the background plasma account for the saturation of the instability 
at moderate amplitudes of the magnetic field.
Previously published MHD simulations {have} assumed a constant cosmic-ray current
{\mkp and no energy or momentum flux in the cosmic rays}, {which excludes a}
backreaction of the generated magnetic field on cosmic rays, and thus the saturation of the 
field amplitude is artificially suppressed. This may explain 
the continued growth of the magnetic field in the MHD simulations. 
{\mkp A strong magnetic-field amplification
to amplitudes $\delta B \gg B_0$ has not been demonstrated yet.}}
\end{abstract}


\section{INTRODUCTION}

{The prime candidate for the sources of Galactic cosmic rays are shell-type
supernova remnants (SNRs), at whose forward shocks a Fermi-type acceleration process
could accelerate particles to PeV-scale energy, although no firm evidence has been
established to date for hadron production in SNRs.
The acceleration arises from pitch-angle scattering 
in the plasma flows that have systematically different velocities upstream and downstream 
of the shock \citep{Bell78}. More detailed studies show that 
the acceleration efficacy and the resulting spectra depend on the orientation angle of 
the large-scale magnetic field and on the amplitude and characteristics of magnetic 
turbulence near the shock \citep[e.g.][]{G+J96,Gia05,Pel06,Mar06}. Particle 
confinement near the shock is supported by self-generated magnetic turbulence \citep{md01}, which is 
likely generated in the upstream region. As the flow convects the turbulence toward and 
through the shock, the turbulent magnetic field structure at and behind the shock 
is shaped by the plasma interactions ahead of it.
Therefore, detailed knowledge of the properties of the turbulence is crucial to 
ascertain all aspects 
of the acceleration processes: transport properties of cosmic rays, the shock 
structure, thermal particle injection and heating processes. 
Of particular 
interest is the question of how efficiently and with what properties 
electromagnetic turbulence is produced by energetic particles at some distance 
upstream of the forward shocks. If the cosmic rays would drive a turbulent 
magnetic field to an amplitude much larger
than the homogeneous interstellar field \citep{lb00,bl01}, particle 
acceleration could be faster and extend to higher energies} {than conventionally 
estimated \citep{lc83}, although it is unclear what energies could be reached
\citep{vladi,ev08}.}

We investigate the properties of magnetic turbulence upstream of the shocks of 
young SNRs. Cosmic rays accelerated at these shocks form a nearly isotropic population
of relativistic particles that drift with the shock velocity relative to the 
upstream plasma. Even though the shock may accelerate particles over a wide range of
energies, the highest-energy particles propagate the farthest from the shock, so at some
distance upstream one finds only particles of the highest energy, which are predominantly
ions. Such a system has been recently studied by \citet[][hereafter denoted as B04]{bell04} 
with a quasi-linear MHD (Magnetohydrodynamics)
approach. Bell noted that, {rather than resonant Alfv\'en waves, the current carried by 
drifting cosmic rays should efficiently} excite non-resonant, nearly purely growing 
($\Re\,\omega=\omega_{\rm R}\simeq 0$) modes of magnetic turbulence. 
{The analytical results were reproduced in studies by 
{\citet{bt05} and \citet{Pel06}, who noted the need to carefully 
account for the return current, and \citet[][hereafter denoted as Z08]{zira08}.}
Calculations by \citet{blasi07} and \citet{rev07} 
using the {linearized} Vlasov equation confirmed the predictions of 
the MHD approximation.} {\citet[][hereafter denoted as B05]{bell05} noted {that} 
a filamentation of the background plasma must
be expected as soon as a small current imbalance occurs, which may eventually
give rise to a filamentation of the cosmic-ray current. 
{MHD simulations {described} by B04, B05 and Z08 indicate} a 
strong magnetic-field amplification following a plasma {filamentation} that {turns
approximately isotropic in the}
non-linear stage. The results presented in those papers do not indicate 
that a parallel plane-wave mode is initially observed.}
  
Here we use Particle-In-Cell {(PIC)} simulations {with a view} 
to explore the properties of the magnetic turbulence produced:
its geometry, non-linear evolution, saturation, and associated particle heating. Our kinetic approach also
allows us to address the backreaction of the magnetic turbulence on cosmic rays. 

{ We describe the simulation setup in \S 2, where we also present
a rationale {for selecting} sets of 
simulations according to the theoretical analyses. In \S 3 the simulation 
results are presented starting with the main three-dimensional experiment,
including a discussion of scaling and parameter 
dependences, and the evolution of the particle spectra. {We conclude with a summary and discussion}
in \S 4. }

\section{SIMULATION SETUP}
\subsection{Simulation Method and Initial Conditions}
Our discussion of the production of magnetic turbulence by the {non-resonant}
streaming instability is based on three-dimensional PIC
experiments. However, performing large-scale simulations in three dimensions
poses serious computational demands. {Simulations of two-dimensional systems {can}
address problems that involve a {wider} range of spatial scales, 
or growth on timescales
long compared with the electron plasma periods, and therefore our three-dimensional experiments
are complemented} by a series of two-dimensional simulations. 
{The two-dimensional runs allow us} 
to study the evolution of magnetic turbulence at timescales for which the
turbulent field structures become larger than the extent of the three-dimensional
simulation {box, and so they} enable us to ascertain the influence of the {
limited size of the simulation box}
on the results of the 3-D experiments. Furthermore, the two-dimensional
runs {permit us to} study how these results depend on the choice of {physical} parameters
of the simulations. 

The code used in this study is a modified parallel version 
of {TRISTAN \citep{bune93}, a} three-dimensional fully relativistic 
electromagnetic Particle-In-Cell code in Cartesian coordinates. 
The modifications to the code include the use of a new first-order 
algorithm for the charge-conserving current deposition developed by \citet{umed03},
and the implementation of a new method of digital filtering of electric currents. 
The filtering method uses the set of 27-point-average isotropic smoothing and compensation 
filters, which attenuate the short-wavelength noise, increase the overall accuracy, 
and at the same time improve agreement with theory at long wavelengths 
\citep[see, e.g.,][]{bird05}. 
The two-dimensional simulations have been performed using the same three-dimensional
code, in which the grid size in one dimension was restricted to contain three cells
only. We thus follow the temporal evolution of all three vector components of particle 
velocities and the electromagnetic fields, and can therefore accurately simulate 
circularly polarized waves as found in the analytical theory,
requiring only that the wavevector lies in the simulation plane. However, a filamentation 
mode may appear somewhat different {in two dimensions.}

In the simulations, an isotropic population of relativistic, monoenergetic cosmic-ray
ions with Lorentz factor $\gamma_{CR}=2$ and number density $N_{CR}$ drifts with 
$v_{sh}=0.3c$
relative to the upstream electron-ion plasma and along a homogeneous magnetic 
field $B_{\parallel 0}$, carrying a current density $j_{CR}=eN_{CR}v_{sh}$. 
The ions of the upstream medium have a thermal distribution
with number density $N_i$, in thermal equilibrium with the electrons. The electron 
population with thermal velocity $v_{e,th}=0.002c$ and density $N_e=N_i+N_{CR}$ 
contains the excess electrons {required} to provide charge-neutrality and also drifts with 
$v_d=v_{sh}N_{CR}/N_e$ with
respect to the upstream ions, {so it provides} a return current $j_{ret}=-eN_ev_d$
to balance the current in cosmic rays. We assume that the return current is entirely 
carried by the electron component of the background plasma because that will most 
quickly react to an electric field induced by charge separation {due to the drifting} 
cosmic-ray component.
The simulations have been performed using {computational grids with} {periodic 
boundary conditions {(see Table 1)}, which keeps
the particle number constant throughout the simulations at a value corresponding to an
average total density of 16 particles per cell}. 
A two-dimensional test simulation with 160 particles per cell has also been run to {verify that
the number of particles used in our studies does not significantly} influence the results.

Cosmic-ray ions drift {in the $-x$-direction,} {antiparallel to the homogeneous 
magnetic field $B_{\parallel 0}$. For the two-dimensional simulations we render the 
$z$-coordinate ignorable,
{so} the direction of the drift is in the simulation plane.}
The electron skindepth $\lambda_{se}=c/\omega_{pe}=7\Delta$, where 
$\omega_{pe}=(N_e e^2/m_e\epsilon_0)^{1/2}$ is the electron plasma frequency
and $\Delta$ is the grid cell size.
{The parameter combination is chosen both to resemble physical conditions in 
young SNRs and, according to {the} quasi-linear calculations by B04, to be} favorable 
for the rapid excitation of purely growing, short-wavelength (as compared with 
cosmic-ray ion gyroradii $r_{CRg}$) wave modes, {\mkp although we typically cannot maintain the
condition $\omega \ll \Omega_i$}. {For a cold ambient
plasma and for wavevectors $k_\parallel$ parallel to $B_{\parallel 0}$, the 
dispersion relation in the unstable wavevector regime 
\mbox{$1 < k_\parallel r_{CRg} < \zeta v_{sh}^2/v_{A}^2$} reads
\begin{equation} 
\omega^2 - v_{A}^2 k^2_\parallel + \zeta v_{sh}^2\, \frac{k_\parallel}{r_{CRg}} = 0, 
\label{eq1}
\end{equation}
for which purely growing solutions are found. The 
maximum growth rate $\gamma_{max}$ at the wavenumber $k_{\parallel max}$ can be determined as}
\begin{equation} 
k_{\parallel max} = \frac{\zeta}{2} \frac{v_{sh}^2}{v_{A}^2}\frac{1}{r_{CRg}} 
\;\;\;\;\;{\rm and}\;\;\;\;\; 
\gamma_{max}      = \frac{\zeta}{2} \frac{v_{sh}^2}{v_{A}}\frac{1}{r_{CRg}}.
\label{e2}
\end{equation}
Here, $r_{CRg}=m_ic\sqrt{\gamma_{CR}^2 -1}/eB_{\parallel 0}$ is the gyroradius of 
the (lowest-energy) cosmic rays,
$v_{A}=[B_{\parallel 0}^2/\mu_0 (N_em_e+N_im_i)]^{1/2}$ is the Alfv\'en velocity, and 
$\zeta=N_{CR}c\sqrt{\gamma_{CR}^2 -1}/N_iv_{sh}$ determines the strength of the cosmic-ray
driving term. From Eq. (1) one can see that the non-resonant modes are strongly
driven when $\zeta v_{sh}^2/v_{A}^2 \gg 1$. For the given parameters, 
this condition depends on the magnetic field strength and the ion-electron mass 
ratio, which also determine  
the wavelength $\lambda_{max}=2\pi/k_{max}$ of the most unstable mode. This wavelength 
must be much smaller than the size of the computational box to enable studies
of the evolution of the turbulence towards larger scales, and at the same time
it must be clearly separated from the physical scales in the background plasma.
Restricted by these computational requirements, for the main three-dimensional experiment (run A)
we assume a {density ratio $N_{CR}/N_i$=1/3 and a 
reduced} ion-electron mass ratio $m_i/m_e=10$, which gives 
the ion skindepth
$\lambda_{si}=25.3\Delta$ and sets the wavelength of the most unstable mode to
$\lambda_{max}=50\Delta$. With this choice we {have further specified} 
$\zeta v_{sh}^2/v_{A}^2\simeq 715$, the Alfv\'enic Mach number $M_A=v_{sh}/v_A=19.1$,
and the ratio $\omega_{pe}/\Omega_e=22.1$, where the electron cyclotron frequency 
$\Omega_e=eB/m_e$, {which corresponds to weakly magnetized conditions of the interstellar 
medium.} The periodic boundary conditions allow us to follow the gyration motion
of cosmic rays, even though the cosmic-ray ion gyroradii, 
$r_{CRg}\simeq 2672\Delta$, are much larger than the size of the simulation box, 
which is 19.8$\times$6.1$\times$6.1 in units of $\lambda_{max}$. 

\subsection{Simulation Parameters}
Table 1 compares the parameters {and main results of all} runs 
in two and three dimensions with those {of the initial} three-dimensional experiment. 
The {follow-up} simulations were mainly used to probe systems with more realistic 
mass ratios up to $m_i/m_e =500$, {better separated spatial scales $\lambda_{si}$ and 
$\lambda_{max}$,} 
{smaller density ratios $N_{CR}/N_i$}, and also to allow the box size to be a higher
multiple of the wavelength of the most unstable mode.
Run A$^{\ast}$ uses the parameters of
run A in {a large two-dimensional box of size 50$\times$18 $\lambda_{max}$ {
and is performed to confirm that the results of two-dimensional runs are similar to those of the 
three-dimensional simulations.} 
Runs B to} E use larger ion-electron mass ratios, {whereas runs G to J are for
smaller density ratios.} 

Note that {in runs B to E} the ion skindepth $\lambda_{si}$ increases with mass ratio as
$(m_i/m_e)^{1/2}$, and thus $\lambda_{max}$ needs
to be readjusted to separate the plasma and turbulence scales. 
{The wavelength of the most unstable mode is thus increased by increasing the 
strength of the homogeneous magnetic field $B_{\parallel 0}$ by approximately 
the same factor $(m_i/m_e)^{1/2}$.}
In this way {the ratio $\lambda_{max}/\lambda_{si}$} and the Alfv\'en Mach number 
$v_{sh}/v_A$ of the cosmic-ray
drift of runs A and A$^{\ast}$ {are} preserved {in those runs}, and also the dynamic range 
{in wavelength,}
as given by $\zeta v_{sh}^2/v_{A}^2$, remains unchanged. However, the maximum
growth rate $\gamma_{max}$ {scales inversely with $(m_i/m_e)^{1/2}$,
and therefore the three-dimensional studies {would} clearly require  an
enormous computational effort, if
more realistic mass ratios were assumed.} We restrict the duration of these simulations 
(runs B and C) to {cover} only the initial stage of the instabilities, and the full
non-linear evolution is investigated by means of two-dimensional simulations (runs D and E). 
{Note that when the ion-electron mass ratio increases,
the upstream plasma becomes more magnetized (the frequency 
ratio $\omega_{pe}/\Omega_e$ {decreases}; {see} Table 1),} and the cosmic-ray gyroradii grow.
For run E, in which $m_i/m_e =500$, these parameters are $\omega_{pe}/\Omega_e=3.3$ and 
$r_{CRg}\simeq 19722\Delta$. Finally, in run F the wavelength of the most unstable mode
has been increased to $\lambda_{max}=150\Delta$ {to allow for a better separation from 
the plasma scales. This is done by increasing the strength of the
regular} magnetic field $B_{\parallel 0}$, whereas the ion-electron mass ratio is {kept} 
the same as in run A. This means that {in this case} the Alfv\'en Mach number 
decreases to $M_A=6.3$ and
also the {width of the unstable wavevector range, here} $\zeta v_{sh}^2/v_{A}^2\simeq 80$, 
is an order of magnitude smaller compared with the other runs. For the parameters of run F,
the ratio $\omega_{pe}/\Omega_e=7.4$ and $r_{CRg}\simeq 890\Delta$. 

{The runs G to J {have been} performed for smaller {values of the}
density ratio $N_{CR}/N_i$ down to 1/30, which 
implies a smaller drift velocity of the electrons in the background plasma. This reduces the impact 
of {an} initial electrostatic, Buneman-type instability, so less heating will occur initially and 
the plasma temperature {should} be lower throughout the simulation. The mass ratio is $m_i/m_e =50$ in 
these simulations {($\lambda_{si}=51.9\Delta$)}, which cover a range in plasma magnetizations 
with parameters
ranging from $M_A=13$ and $\lambda_{max}=500\Delta$ to $M_A=65$ and $\lambda_{max}=100\Delta$.}
In all simulations the time step $\delta t=0.0715/\omega_{pe}$.

\section{RESULTS}

\subsection{Three-dimensional Experiment}
The temporal evolution of the magnetic and electric fields and the particle kinetic-energy
densities in the three-dimensional simulation (run A) is shown in Fig. 1. 
An initial (up to $t\sim\gamma_{max}^{-1}$) fast growth of the turbulent 
fields is caused by a Buneman-type beam instability
between drifting electrons and the ambient ions, which leads to 
plasma heating until the electron thermal velocities become comparable
to the electron drift speed $v_d$ {(recall that the electrons in the background plasma must slowly
drift to initially compensate the cosmic-ray current)}.  found in the analytical theory
In real SNRs the density ratio {$N_e/N_{CR}$ is likely
much larger than the sonic} Mach number, and therefore the electron drift speed
is smaller than their thermal velocity. The initial instability is thus solely
a consequence of the initial conditions in our simulations, in particular the
initial temperature of the background medium {and the density contrast between cosmic rays and
the background plasma.} 

{At $t\sim 2.5\,\gamma_{max}^{-1}$ wave modes start to emerge that are
excited by the cosmic-ray ions streaming in the upstream plasma, and turbulent
magnetic field is seen} mainly in the components transverse 
to the cosmic-ray drift direction. A {plane-wave Fourier analysis is shown
in Fig.~\ref{fBy} for one component of the magnetic field and in Fig.~\ref{fNe} for the density
of the background electrons. {All Fourier power spectra are averages 
over the simulation box to reduce the noise.} Note that for a plane wave of wavelength $\lambda$ 
propagating at an angle $\theta$ to the drift ($-x$) direction, the one-dimensional
Fourier spectra will reveal a signal at $\lambda_\parallel=\lambda_x=\lambda/\cos\theta$
and at $\lambda_\perp=\lambda_z=\lambda/\sin\theta$, respectively. Assuming a
plane wave as in the analytical calculations of B04 we therefore conclude} that the dominant
wave mode is oblique at an angle of about $\theta \approx 80^\circ$ 
to the $x$-direction, and has the perpendicular wavelength 
component $\lambda_\perp \simeq 50\Delta$, which is numerically similar to $\lambda_{max}$.
The character of the mode is thus different than {predicted by the quasi-linear analysis 
(B04, B05), which indicated that {\mkp initially} the most rapid growth should occur} for
wavenumbers parallel to $B_{\parallel 0}$ (see also Eq. 1). 
{Here we observe a modified}
filamentation instability, which appears to be faster than the non-resonant parallel 
modes {\mkp and was also observed in MHD simulations (B04, B05, Z08 \citet{rev08}) to dominate in
the non-linear phase.} {\mkpa At that time, when $\delta B > B_{\parallel 0}$ is already established, 
the field structures in the MHD studies have the same spatial scales parallel and perpendicular 
to the cosmic-ray drift direction, as evident from Figure 3 in Z08. As we describe in 
more detail below, the same is true for the magnetic structures in our simulations, which at that point more
resemble cavities and isolated peaks.}  
The appearance of an oblique filamentary mode is in line
with earlier results for electron beams which show that the strongest growth is typically 
observed not for {the perpendicular case} ($\vec k \perp \vec v_{sh}$), but at a finite angle
\citep{bret05, dieck08}, 
because there is a cumulative effect of instabilities that operate in parallel 
\citep{laz07}, which naturally is not captured in 
linear analytical analysis. The appearance of magnetic power spectra is then 
further complicated by the fact that filamentation is not exactly a 
transverse {mode {in the sense that $\vec k\cdot\vec E\neq 0$}} \citep{bret07}.
As seen in Fig.~\ref{fBy}, the {scale in $\lambda_\perp$ of the dominant feature in the spectrum}
grows as the turbulence develops, but the parallel scales 
($\lambda_\parallel=\lambda_x \simeq 250\Delta$) remain roughly constant, so after about 
$20\,\gamma_{max}^{-1}$ the structures have similar size parallel and perpendicular
to the drift. A comparison with Fig.~\ref{energy} reveals that at those late times 
$\delta B_\perp$ is of the {same} order {as} $B_{\parallel 0}$, so the 
system becomes magnetically nearly isotropic.

The spatial structure of the turbulent magnetic field and {the density of} the ambient
plasma are shown in Figs.~\ref{ne-ni10}, \ref{nib10}, and \ref{nibXZ10} 
for $t\approx 10\,\gamma_{max}^{-1}$, and found in the analytical theory
Fig.~\ref{multi} presents snapshots of the {temporal} evolution of these structures in the plane
perpendicular to the cosmic-ray drift direction. 
{To be noted from Fig.~\ref{ne-ni10} is that the density variations of the ambient electrons 
and ions are nearly cospatial, apart from some variations on small scales that arise largely 
from statistical
fluctuations, so it is sufficient to only show the ion density in 
Figs.~\ref{nib10}, \ref{nibXZ10}, and \ref{multi}.}
The background plasma forms filamentary 
structures of plasma voids surrounded by regions of enhanced plasma density. 
At any given time,
the {thickness of these structures, or their separation,}
corresponds to the $\lambda_\perp$ component of the dominant wave mode,
and the correlation length along the cosmic-ray streaming direction is  
equal to $\lambda_\parallel$. 
{The cosmic-ray distribution remains 
homogeneous throughout the duration of the numerical experiment, but the}
voids become completely depleted of ambient ions {and electrons} {except for the 
excess electrons which neutralize the charge in cosmic-ray ions}. 
{The currents carried by cosmic rays are {therefore} no longer neutralized, and} 
{a strong net current {flows} inside the filamentary  
ambient-plasma voids, {resulting in} magnetic-field lines circling around 
the center of the cavities according to Amp\`ere's law.  
{Such filamented current structures with azimuthal magnetic fields 
and radial electric fields are also observed in other simulations \citep{nishi06}.}

The perpendicular magnetic field is thus concentrated around regions of low background 
plasma density, but constant cosmic-ray density.}  
In the MHD {description}, the formation of cavities in the plasma 
is due to the $\vec{j}_{ret}\times\vec{B}$ force, which accelerates the ambient
plasma away from the center of the voids, thus causing the cavities to expand
\citep[B04, B05,][]{mil06}. In our kinetic simulations, {mainly the electrons 
are subject to this force, since they drift and thus} contribute to $\vec{j}_{ret}$.
Ambient-ion motion away from the cavities, {and a restoring influence on the electrons,}
is caused by an electrostatic charge-separation field, which {is} stronger than the original 
$\vec{j}_{ret}\times\vec{B}$ force for non-relativistic drifts.

\subsection{Theoretical Analysis of Current Filament Structure}
{The formation of the low-density filaments in the ambient plasma can be understood 
with a simple toy model. We use cylindrical geometry with azimuthal symmetry and assume 
the cosmic rays to drift homogeneously along the  $z$-axis, thus providing a current density
$j_{\rm CR}={j_{\rm CR,z}}=e\,N_{\rm CR}\,v_{sh}$. As in our simulations, the ions in the background 
plasma are at rest and homogeneously distributed with density $N_{i}$, whereas the background 
electrons are initially homogeneously distributed with density $N_{0e}$ and 
drift along the $z$-axis with velocity $v_{\rm d}=v_{\rm d,z}=v_{\rm sh}\,N_{\rm CR}/N_{\rm 0e}$ to 
cancel the cosmic-ray current and charge.
Now suppose a small displacement of electrons occurs at some location, so that
\begin{equation}
\delta N_e (r)=-N_{0e}\,\eta\,r_1\,\left[\delta(r-r_1)
-{{r_1}\over {r_2}}\,\delta(r-r_2)\right]
\label{mod1}
\end{equation}
where we assume for simplicity $r_2 > r_1$ and introduce $\eta >0$ as a small perturbation parameter.
The displacement breaks the current balance, so that 
an excess current exists with
\begin{equation}
\delta j_z (r)=j_{\rm CR}\,\eta\,r_1\,\left[\delta(r-r_1)
-{{r_1}\over {r_2}}\,\delta(r-r_2)\right]
\label{mod2}
\end{equation}
This current creates an azimuthal magnetic field that is determined by Amp\`ere's law.
The field is non-vanishing only between $r_1$ and $r_2$ and has the amplitude
\begin{equation}
B_\phi (r)=\mu_0\,j_{\rm CR}\,\eta\,{{r_1^2}\over r}\qquad r_1 < r < r_2
\label{mod3}
\end{equation}
The small displacement of electrons also implies a violation of charge balance, leading to a
radial electric field of amplitude
\begin{equation}
E_r (r)={{j_{\rm CR}}\over {\epsilon_0\,v_{\rm d}}}\,\eta \,{{r_1^2}\over r}\qquad r_1 < r < r_2
\label{mod4}
\end{equation}
The electric force on the electrons is therefore a factor $c^2/v_{\rm d}^2 \gg 1$ larger than the
average magnetic force, so they are accelerated inward. 
The ambient ions see on average only the electric force, 
because at least initially they do not drift, which accelerates them outward, 
albeit with an acceleration much 
smaller than that of the electrons. The cosmic rays are virtually unaffected on account of 
their large random velocities, which implies that they see any acceleration only for a very short time. 
The dominant electric field prevents a separation of ambient ions and electrons, and in fact we see
in Fig.~\ref{ne-ni10} that these two particle species are nearly cospatial. The
ambient ions and electrons readjust {their {spatial} distributions} until charge balance is 
re-established, but as a bulk they 
move slightly outward, so the currents are not balanced. Consequently, a current-density profile is 
established that resembles that described in Eq.~\ref{mod2}, although the parameters 
$\eta$, $r_1$, and $r_2$
will not be the same as before. Nevertheless, a magnetic field similar to that in Eq.~\ref{mod3} results,
that is not compensated by an electric field. {The ${\bf v}\times{\bf B}$ force}
will pull electrons outward and with them the ambient ions,
because they are electrostatically bound to the electrons. Reusing Eq.~\ref{mod2} with modified 
parameters $\eta^\prime$, $r_1^\prime$ {and with $\omega_{\rm pi}$ as the ion plasma frequency,}
the radial acceleration of the ambient plasma can be written as 
\begin{equation}
a_r (r)=\omega_{\rm pi}^2\,r_1^\prime\,\eta^\prime\,{{v_{\rm d}^2}\over {c^2}}\,{{r_1^\prime}\over r}
=\omega_{\rm pi}^2\,r_1^\prime\,\eta^\prime\,{{N_{CR}^2}\over {N_e^2}}
\,{{v_{sh}^2}\over {c^2}}\,{{r_1^\prime}\over r}\label{mod5}
\end{equation}
so the growth time can be estimated as
\begin{equation}
t_{\rm filament}\simeq \sqrt{{r_1^\prime}\over {a_r}}={1\over {\sqrt{\eta^\prime}\,\omega_{\rm pi}}}\,
{{N_{e}}\over {N_{CR}}}\,{c\over {v_{sh}}},
\label{mod6}
\end{equation}
{which recovers the scaling of Eq. 2 {and is identical to the result of B05 for the
filamentary mode.} {It also shows the same scaling as the expected growth rate of the parallel
plane-wave mode (see Eq.~\ref{e2}).}
For} efficiently accelerating SNRs $t_{\rm filament}$ could be of the order of hours or days, 
if $\eta^\prime$ were close to unity.}

{If the initial displacement of the electrons were inward ($\eta < 0$ in Eq.~\ref{mod1}), 
an inward transport of ambient plasma would result. However, the electrostatic restoring force
(Eq. \ref{mod4}) would be stronger on account of the $1/r$-profile, and therefore a weaker current 
imbalance 
would result after the re-establishment of charge balance in the background plasma, thus imposing a 
preference for outward displacement of the ambient plasma and the creation of plasma voids.} 
  
\subsection{Non-linear Evolution of Current Filaments}
The expansion of the plasma cavities is visible in Figure \ref{multi}, and also as the 
evolution toward larger scales in the power spectra shown in Figs.~\ref{fBy} and \ref{fNe}. 
The growth of the cavities and the subsequent merging of the adjacent plasma voids
lead to a compression of the plasma between cavities and also to an amplification
of the magnetic field. Because the magnetic field has a preferred orientation around
each cavity, the magnetic field lines {may} cancel each other in the space between the
voids (Fig. \ref{nib10}). The turbulence is almost entirely magnetic and there is no 
large-scale turbulent
electric field structure accompanying the filamentary distribution of ambient
plasma and perpendicular magnetic field. The electric fields result from  
thermal plasma motions, the level of which can be estimated from Figure \ref{ne-ni10}. 

Compared with Bell's MHD studies, the growth of the magnetic 
perturbations in our kinetic modeling is much slower than calculated
{for the parallel planar mode}. The initial 
growth rate {of the perpendicular-field turbulence is only 
$\sim 0.2\,\gamma_{max}$ {prior} to $t\gamma_{max}\approx 7$}, and becomes smaller during
the later evolution, possibly after nearby plasma cavities have started to merge. {
In the sense of the toy model discussed in {\S 3.2} (Eq.~\ref{mod6}),
the perturbation parameter must be small, $\eta'\le 0.1$, or neighboring filaments partially cancel the 
magnetic field between them.} 
Moreover, the amplitude of the turbulent component never 
considerably exceeds the amplitude of the regular field.
{Already after about $20 \gamma_{max}^{-1}$, 
when $\delta B_\perp \approx B_{\parallel 0}$,
the growth of the turbulence starts to saturate. 
As shown in Figures \ref{energy}, \ref{fBy}, and \ref{multi},
the initial {filamentary} structure of plasma cavities and {surrounding}
magnetic fields becomes disrupted at this stage, and 
the turbulence is nearly isotropic and highly non-linear. In the
simulation frame, which is the {initial rest frame of the} ambient ions, 
the turbulent field structures {start to
move {as a consequence of being} embedded in the background plasma which itself starts
to drift (see \S~\ref{sec3.3}).} 
Merging of the magnetic structures leads to
further amplification of the magnetic field through compression, but at very
slow rate. {The peak amplitude of the average turbulent field,
$\delta B_{\perp}^{max} \simeq 3.5 B_{\parallel 0}$, is
reached at $t\gamma_{max}\approx 40$,} after which the field starts to dissipate. 
This is related
to the {increasing length scales} {and dissipation} of the structures in the
ambient-plasma density as evident from
Figure~\ref{fNe}. The electron and ion distributions become homogeneous, but
still evolve nearly cospatially, so that currents produced in this process are weak
and thus the gradual decay of the magnetic field is slow.

{Also clearly visible in Figure~\ref{fNe} is that at 
$t > 25 \gamma_{max}^{-1}$ the dominant structures become
larger than the size of our simulation box. Using two-dimensional simulations 
performed on a large computational grid (run A$^{\ast}$), we have verified that
the long-time evolution of the magnetic turbulence and the plasma density fluctuations}
in the three-dimensional experiment are not considerably influenced by the {
size of the simulation box}. 
Figure \ref{energy2-D} shows the temporal evolution of the energy
densities in particles and fields for run A$^{\ast}$, and Figure \ref{fBy2-D} 
presents the power spectrum evolution of the perpendicular magnetic field 
component. {To be noted from the figures is} that 
the initial growth of the perpendicular magnetic field, associated
in part with the Buneman-type instability, leads to the higher initial 
amplitude of the transverse field component in the two-dimensional simulation,
and thus  $\delta B_\perp$ reaches the amplitude of the regular field faster
compared with run A. Otherwise, the results of the three-dimensional experiment
are qualitatively very well reproduced in the two-dimensional run. {At about 
$t\gamma_{max}\approx 35$ the peak amplitude of the average turbulent field is reached,
and its value, $\delta B_{\perp}^{max} \simeq 3.1 B_{\parallel 0}$, is} very close to that 
{obtained in} the three-dimensional simulation. Also, the wavelengths 
($\lambda_\perp$, $\lambda_\parallel$) of the dominant wave mode of the magnetic-field
turbulence are numerically similar to those in run A.} The dominant
turbulent field structures during the late-time evolution are well contained 
inside the simulation box (Fig. \ref{fBy2-D}), so that the eventual dissipation 
of the magnetic field is not affected by the boundary conditions.    

\subsection{Scaling and Parameter Dependence}
The question arises to what extent the parameter choice in our simulations influences 
the plasma dynamics, in particular the turbulence growth rates.
We have used the zero-temperature kinetic calculations of \citet{blasi07} to
verify that our choice of the reduced ion-electron mass ratio $m_i/m_e=10$ and a monoenergetic 
cosmic-ray spectrum with particle Lorentz factor $\gamma_{\rm CR}=2$ has no impact on the 
{growth rate and spatial scale of the} instability. 
{For that purpose} 
we have re-evaluated their approximations for $\omega\ll\Omega_i^\ast$, 
the nonrelativistic ion gyrofrequency, and {as the only impact of our parameter choice} 
{we} find an additional term in the dispersion relation that scales with the mass ratio.
In the notation of Eq. 2 in \citet{blasi07}, the dispersion relation would read
\begin{equation}
v_A\,k^2 -{{m_e}\over {m_i}}\,{{N_{\rm CR}}\over {N_i}}\,v_{sh}^2\,k^2
=\omega^2 \mp{{k\,v_{sh}\,\Omega_i^\ast\,N_{\rm CR}}\over {N_i}}\,\left[1\pm I_1(k)\pm\imath\,I_2 (k)\right]
\label{A1}
\end{equation}
For a realistic mass ratio and the parameters used by \citet{blasi07}, the additional term is 
negligible, but for the standard parameters in our simulations it is not. In fact, the left-hand
side of Eq.~\ref{A1} is negative for all $k$, and hence for one polarization we find a nearly purely
growing solution with a growth rate slightly larger than for 
$m_i/m_e=1836$ for all $k\,r_{CRg} \gg 1$
because the functions 
$I_1 \ll 1$ and $I_2 \ll 1$ for those wavenumbers.
These calculations suggest that the {small growth rate}
of the non-resonant streaming instability 
in our simulations is not caused by the reduced ion-electron mass ratio, {in line with the
results of additional 2-D simulations using more realistic mass 
ratios (runs B-E), whose results are described below.}

Figure \ref{energy500} shows the temporal evolution of the energy density in electromagnetic fields and 
particles for a two-dimensional run with $m_i/m_e=500$ (run E). One can note from the figure and also 
from Table 1 that the fundamental properties of the magnetic turbulence observed in the three-dimensional 
experiment (run A) are also seen in the additional simulations {for $m_i/m_e=40,100,150$ and $500$}.
In particular, for all mass ratios the dominant wave mode, which is the oblique 
filamentary mode, has approximately the {same growth rate, much slower than predicted {
by the quasi-linear estimates,} and a wavevector with the same inclination to the drift 
direction $\theta\approx 70^\circ$}. The two-dimensional simulations {for high 
ion-electron mass ratios} (runs D and E) 
also show that the {saturation level of the magnetic turbulence is} similar 
to that observed in the three-dimensional experiment with $m_i/m_e=10$.    
 
{Finite-temperature effects in background plasma can limit the growth rate of 
the non-resonant instability \citep{rev06,rev07}}. Since our {simulations} do not exactly 
reproduce the {analytical estimates for the} zero-temperature limit, we used the 
analytical scalings of \citet{rev07}
to assess the role of the thermal effects. The initial heating of the ambient-particle
populations associated with, e.g., the Buneman instability leads to ion temperatures of the order 
$\Theta=k_BT/m_ic^2\approx 10^{-4}-10^{-3}$. {At these temperatures the growth rates for
the most unstable wave modes should be reduced as 
$\gamma_{max}(\Theta=10^{-4})\simeq 0.9\, \gamma_{max}(\Theta=0)$ and
$\gamma_{max}(\Theta=10^{-3})\simeq 0.5\, \gamma_{max}(\Theta=0)$.\footnote{We are indebted to B. 
Reville for providing these estimates to us.} 
This means that thermal effects alone cannot {explain why in our simulations
a filamentation mode with a growth rate of $0.2\, \gamma_{max}(\Theta=0)$ is observed, but not
the parallel non-resonant mode.}

For the two-dimensional run F with $m_i/m_e=10$ we have 
increased the strength of the regular magnetic field component $B_{\parallel 0}$. In that case the 
turbulence scales ($\lambda_{max}$) are much better separated from the plasma scales; in fact,
$\lambda_{max}$ is a factor of three larger than in the runs A and A$^{\star}$, whereas
the electron and ion skindepths are the same as before.} In this way we also
probe the production of magnetic turbulence in more magnetized upstream plasma, in which the shock 
is weaker {with an} Alfv\'enic Mach number $M_A=6.3$. The results are presented in Figures \ref{energy150} 
and \ref{fBy150} {and summarized in Table~1, where the reader will note 
that the growth of the turbulence is faster than that for stronger shocks with a maximum
rate 0.35~$ \gamma_{max}$. However, the saturation level is essentially the same as
in all the other simulations. Furthermore, the faster evolution allows us to better follow the 
late-time behavior of the system, and we clearly observe the turbulent magnetic field to slowly decay until
at the end of the simulations the energy density in the turbulent magnetic field is within a factor 2 of that
in the homogeneous magnetic field. We conclude that there is no evidence for significant
magnetic field amplification, {although we} do observe a small change in the dominant oblique filamentary 
mode, which has a wavevector that is slightly better aligned with the drift direction
($\theta=\angle(\vec k, \vec v_s) \approx 53^\circ$), {which appears to be a general trend
when $\lambda_{max}$ is expected to be a higher multiple of the ion skindepth.}}

{We also performed a number of simulations with reduced $N_{CR}/N_i$, for which the expected growth rate of 
the non-resonant instability {decreases} according to the quasi-linear estimates. A smaller cosmic-ray
current implies that the electrons in the background plasma can compensate that current with a smaller drift
velocity, thus lessening the impact of the initial electrostatic Buneman-type instability and the associated 
heating. The runs G to J therefore involve lower plasma temperatures throughout the duration of each simulation.
The growth of the magnetic turbulence becomes very slow {in these simulations, which use the same 
$\delta t \ll \omega_{pe}^{-1}$ as runs A-F, and therefore} in run J with $N_{CR}/N_i=1/30$ we follow 
only the initial evolution. The runs G-I are for $N_{CR}/N_i=1/10$ and three different magnetizations,
implying that the {expected} turbulence scales, $\lambda_{max}$, are between 2 times and 10 times
the ion skindepth. 
The mass ratio is $m_i/m_e=50$ for all runs with small density ratios. As in the case of runs A-F, the mean
amplitude of the turbulent magnetic field peaks at a few times the homogeneous field strength and we observe
an oblique filamentation mode, not the parallel mode expected from quasi-linear theory.}

\subsection{The Evolution of the Particle Spectra}\label{sec3.3}

{In all runs a} random sample of {0.1\% - 0.3\%} of each particle species 
(electron, ion, cosmic ray), {corresponding to about $10^6$ particles,} {is}
selected every few hundred timesteps to observe the evolution of the velocity distributions.
The average particle velocity is the particles' drift velocity and is plotted for all three
species in Fig. \ref{f-drift} {for run A}. To be noted from the figure is the disappearance of a 
relative drift between
the cosmic rays and the background ions and electrons: after approximately $40\, \gamma_{max}^{-1}$ all particles 
drift with approximately the same bulk velocity of
about $0.14\,c$. 
At the same time the turbulent magnetic field reaches its peak energy
density, as is shown in Fig.~\ref{energy}. The same basic behavior is seen in the other simulations
as well. This finding is in line with the notion that the cosmic-ray
drift relative to the background plasma drives the turbulence: as the relative velocity between 
cosmic rays and plasma decreases, so does the source of the non-resonant
streaming instability. {\mkp We have performed a test simulation (run I$^\ast$ in Table 1), in which we do 
not allow any acceleration of cosmic rays by increasing their particle mass
by a factor $2\times10^8$ and leaving all other parameters as in run I. Indeed, the cosmic-ray
drift remains constant and the magnetic field grows to a bit more than twice the amplitude
as compared with run I. The background plasma would still experience a bulk acceleration, 
but now the magnetic-field growth terminates only when the plasma has assumed the full original
cosmic-ray drift speed rather than about half of it as in all the other simulations, 
thus delaying the saturation.}  

{The bulk
acceleration of the background plasma plays a key role in shaping cosmic-ray modified shocks, 
although in realistic SNRs the process must be expected to operate on a timescale similar to 
that on which the local conditions change on account of the approaching shock.} {In a steady state
the upstream plasma will assume a bulk velocity that is essentially determined by} the local density 
of cosmic rays. In any case one expects a continuous change in the plasma flow velocity, which permits
the overall compression ratio between far upstream and far downstream to be much larger than
the limit given by the Rankine-Hugoniot jump conditions \citep[e.g.][]{vladi}. Our simulations
detail how the changes in the bulk-flow properties relate to the average amplitude of the magnetic field.

Besides the buildup of magnetic turbulence and the bulk
acceleration of the background plasma we can also study the heating of the plasma, which is of particular
interest for models of cosmic-ray acceleration because it limits the sonic Mach number of the shock
and hence could reduce the acceleration efficiency. 
After a Lorentz
transformation into the bulk-flow frame {of a given plasma component} (see Appendix~\ref{spectra} for details), 
measurements of the azimuth-integrated particle distribution 
can reveal anisotropies and spectral evolution. {Such an analysis shows that}
all particle species, 
cosmic rays as well as the background ions and electrons, remain moderately isotropic in their 
instantaneous bulk-flow rest frames, but a certain stretching of the distributions along the
drift direction is clearly seen. Therefore, care must be exercised in deriving and interpreting the
momentum spectra of the particles. 

The cosmic rays start the simulation with an isotropic and monoenergetic distribution {with
Lorentz factor $\gamma_{\rm CR}=2$.} During 
the simulation isotropy is maintained to within 0.5\%, and the momentum distribution of 
the cosmic rays continuously broadens, but remains approximately {a Gaussian centered
on $mc\sqrt{\gamma_{\rm CR}^2 -1}$} with a mode that
does not change appreciably.

The momentum distributions of the background ions and electrons intermittently
develop a high-energy tail, at least part of which can be attributed to anisotropy because
the particles in the high-energy tail predominantly move along the drift direction.
A more detailed inspection, however, reveals that {most of the} anisotropies arise from the 
initial electrostatic Buneman-type instability, because they appear long before the non-resonant 
magnetic instability sets in. The initial electrostatic
instability involves essentially only electric fields parallel to the drift direction which significantly
stretch the distributions of slow particles, which is why the background ions are most strongly affected
with peak anisotropies of 20\%.
The Lorentz force contributed by the perpendicular components of the magnetic fields provides 
re-isotropization which is faster for electrons than for ions on account of the difference in gyrofrequency,
but even in the case of the ions the anisotropy is down to $\lesssim\,$3\% 
{after about $t\,\gamma_{max}\simeq 15$ in the {main} three-dimensional simulation.} 

As outlined in appendix~\ref{spectra}, we can split the kinetic energy density of particles into the
components associated with the bulk motion and the random motion. We denote increases in the 
random energy density as heating, but the reader should note that this does not {necessarily} 
imply a thermalization. {Figure~\ref{energy} shows the temporal evolution of the random energy density
for all three particle species in run A.}
Substantial heating of both the electrons and the ions
in the background plasma is observed early in the simulation
as a result of the initial electrostatic instability. Further strong heating occurs after
$t\,\gamma_{\rm max}\simeq 20$, when the random energy density of all background plasma species is always
higher than their bulk energy density. The mean random kinetic energy per particle, or ``temperature,'' 
is typically different for the electron
and ions in the background plasma. After the initial Buneman-type instability the electrons are hotter,
but roughly at the time the filamentation instability has taken over and turned non-linear, the ions are
strongly heated {and they eventually} become hotter than the electrons.
{These} results on the anisotropy and heating
{are seen in} the main three-dimensional simulation, {but 
qualitatively the same behavior is observed also in two-dimensional runs.}

If the observed heating of the background plasma were real, then the upstream medium of an SNR forward
shock would be so hot that {the forward}
shock itself could be only very weak, if it existed at all. There are 
reasons to assume that our simulations for technical reasons overestimate the heating, {and
therefore we cannot make {firm} statements on the temperature of a real plasma upstream of 
SNR shocks with efficient particle acceleration.} One reason is
that we must limit the number of simulated particles to simulate the long-term evolution in large 
simulation boxes. Statistical fluctuations then give rise to small-scale electric fields that heat the 
plasma. {A short test simulation
with large particle number (160 per cell in total) was performed and indeed the plasma temperature was
observed to be somewhat reduced, although the ions eventually reached a temperature similar to that observed for 
smaller particle number. As statistical fluctuations are suppressed only with $\sqrt{N}$, simulations
with low density ratio and a sufficiently large number of particles per cell are prohibitively expensive, even 
in two dimensions. Other computational techniques like $\delta f$-PIC simulations \citep{sydora}
may be better suited for that purpose.}

Also, a substantial part of the initial heating and small-scale electric fields is related to the initial
electrostatic Buneman-type instability, the effect of which depends on the drift velocity of the background
electrons. Because the electrons carry the return current that compensates the cosmic-ray current, their 
drift velocity is $v_d=v_{sh}N_{CR}/N_e$. We must keep the shock speed $v_{sh}$ (the
cosmic-ray drift velocity) high to enable a sufficiently fast growth of the non-resonant streaming
instability. We can vary the density ratio $N_{CR}/N_e$ somewhat using particle splitting, but that also
adversely affects both the growth rate and the wavelength of the non-resonant streaming
instability. We have run {three test simulations with $N_{CR}/N_i=1/10$ and one
with $N_{CR}/N_i=1/30$ to further explore this issue. We find that the plasma temperature, or more precisely
the average random kinetic energy per particle, is indeed significantly reduced by about an order of magnitude 
throughout the simulation. Nevertheless, the buildup of magnetic turbulence is unchanged compared with the earlier
experiments. The plasma temperature is still {unrealistically high}, probably on account of heating 
in small-scale electric fields that arise from statistical charge-density fluctuations. }
  
\section{Summary and Discussion}
\subsection{Summary of Our Results}
The {generation} of magnetic-field turbulence by cosmic-ray ions
drifting upstream of SNR shocks has been studied using {two- and three-dimensional}
PIC simulations {for a variety of parameters}.
Turbulent field is indeed generated in this process, but
the growth of magnetic turbulence is 
slower than estimated {in the literature, and the turbulence is
of a different nature: we observe a modified filamentation of the ambient plasma, 
but not the cosmic rays, in contrast to the parallel wave found in quasi-linear calculations.}
{\mkp The filamentation and formation of cavities in the background plasma was also observed in 
recent MHD simulations.}
 
The amplitude of the field
perturbations saturates at approximately the amplitude of the homogeneous 
upstream field. The energy density in the turbulent field is also
always much smaller than the plasma kinetic energy density. This suggests that
the efficiency of magnetic-field generation through this mechanism may not be sufficient
to account for the {strong magnetic-field amplification invoked for} some young SNRs and also
leaves open the question {whether or not diffusive particle acceleration 
at SNR shocks can} produce particles with energies beyond 
the ``knee'' in the cosmic-ray spectrum. 

{The backreaction of the magnetic turbulence on the particles leads to an 
alignment of the bulk flow velocities of the cosmic rays and the background medium in all our simulations.}
This is
precisely what is making up a cosmic-ray modified shock: the upstream flow speed is continuously changed 
by the cosmic rays, so the compression ratio of the actual shock of the thermal plasma is moderate, whereas
the overall compression ratio can be large. The new and surprising result of our simulations is that we 
accomplish this without significant magnetic-field amplification and with instability modes different from
those invoked for the purpose in the recent literature.

\subsection{Comparison with Published MHD Simulations}

{How can we understand the differences between the results of the
analytical solutions, the MHD simulations, and our PIC {study}? First of all,
{\mkp not too much} is different:
the filamentation mode observed in our simulations is similar to the cylindrical {filamentary} mode that was
analytically described in B05 {\mkp and {\mkpa in its evolved, then isotropic form}
observed in the MHD simulations of B04, B05, Z08, \citet{rev08}}.
The growth rate, or expansion rate, of the filaments
is somewhat smaller than expected for isolated filaments, probably because neighboring filaments produce
between them magnetic fields of opposite orientation, thus cancelling part of the magnetic field that
one would calculate for an isolated current filament. We do not observe a {filamentation} of the
cosmic rays, probably because throughout the simulations
the cosmic-ray Larmor radius, {$r_{CRg}$},
remains larger than the spatial scale of the magnetic turbulence. As we observe the
magnetic-field growth to saturate and the cosmic-ray drift relative to the background plasma to disappear, 
there is no significant cosmic-ray current remaining that could initiate a further
amplification of the magnetic field at later times, beyond the termination of our simulations.

The MHD simulations of B04, B05, and Z08 show the magnetic-field growth to
an amplitude much higher than the initial homogeneous field. 
{However, they assume the cosmic-ray current to be constant throughout the simulations.}
In our {study} the cosmic-ray current changes and is 
strongly reduced in the non-linear phase, when the magnetic-field growth saturates. Because 
{in the MHD simulations the cosmic-ray current is {constant} in time and uniform in space, these 
simulations consequently cannot} capture this backreaction. 
{\mkp In the test run I$^\ast$ (see \S\ref{sec3.3} and Table 1), in which the 
backreaction of cosmic rays has been excluded by increasing cosmic-ray particle 
mass, we observe the peak in the 
average magnetic-field strength more than a factor 2
higher than with cosmic-ray backreaction (run I). In this test simulation
the saturation of the field amplification arises from the bulk acceleration of the ambient plasma
which is still permitted. In the MHD simulations, it is unclear to what extent momentum is transferred 
to the background plasma, but in any case that process is likely suppressed because the energy and 
momentum flux of the cosmic rays is not accounted for.}

{B04, B05, and Z08 describe the magnetic-field structure in the MHD simulations} at various stages
of the evolution. The non-linear} turbulence appears fairly isotropic and bears some resemblance to 
the structure shown in Figs. \ref{nib10} and \ref{multi}, although the 
{MHD results reveal} no structures visibly 
extended in the drift direction. The MHD simulations show much more structure on the smallest scales, 
most of which are prominent at late 
stages when the turbulence in our PIC simulations
has already saturated. The one-dimensional spectra of the perpendicular magnetic field in the MHD simulation
of Z08 indicate strong growth on all spatial scales {after} the turbulent field 
{becomes} stronger 
than about 10\% of the homogeneous field. From that time on the velocity fluctuations 
in the background plasma are also of the same order as, if not stronger than, the adiabatic sound speed, 
indicating that weak shocks can be formed. In our simulations shocks would be resolved, which may 
explain why the MHD simulations show more structures on the smallest scales. 

The discussions in B04, B05, and Z08 do not indicate that
a parallel planar mode is indeed initially observed in the MHD simulations.
Also, neither magnetic-field spectra in $k_\perp$ nor spectra of the parallel component of the 
magnetic field are presented. Our discussion of the differences between
the results of {these} MHD simulations and our PIC {study} must therefore
remain limited in scope. We stress, however, that the saturation of the turbulence growth in our simulations
arises from the backreaction {of the magnetic field} on the bulk velocities of cosmic rays and background plasma. 
The higher {magnetic-field} saturation level in the MHD simulations is most likely a direct 
result of {the assumption of} a constant cosmic-ray current. {\mkp A strong magnetic-field amplification
to amplitudes $\delta B \gg B_0$ has yet to be demonstrated.}

\subsection{Comparison with Analytical Calculations}
{\mkp The turbulence observed in our simulations reflects analytical results
for filamentation modes, e.g., the derivation by B05. A discrepancy only exists as
far as the parallel plane-wave mode is concerned.}
All published analytical calculations agree that in quasi-linear treatment a parallel, 
purely growing mode should dominate. What might be the reasons why we do not see this mode in 
our PIC simulations? We have already determined that the plasma temperature is probably not 
responsible for this discrepancy. One possibility is that the mode exists initially,
but is invisible in the noise, and changes {its} character quickly so that we observe the filamentation, which
was already described in B05. {In fact, the parallel plane-wave mode and the filamentation of 
the ambient plasma are expected to show a similar growth rate.} This interpretation 
would be in line with the magnetic-field structure observed in the MHD simulations before the turbulent magnetic 
field reaches the amplitude of the homogeneous field, $B_{\parallel 0}$, which also does not clearly show a 
parallel mode.

{\mkp The most likely reason} is that our choice of parameters may not reflect one of the assumptions made in
the analytical treatments, {\mkp namely} that the frequency of the perturbations be much 
smaller than the ion gyrofrequency \citep[e.g.][]{rev07}, whereas we typically have to 
choose parameters for which the theoretically expected growth
rate is similar to the ion gyrofrequency, i.e. $\Im\omega_{}\approx \Omega_i$.  
According to the quasi-linear results,
\begin{equation}
{{\Im\omega}\over {\Omega_i}} \approx {{v_{sh}\,N_{CR}}\over {2\,V_A\,N_i}}\ ,
\label{limit1}
\end{equation}
implying that if either the shock speed or the cosmic-ray density is too high, the assumption of a small 
frequency is no longer justified, and the character of the instability may be different. However,
the observed growth rate 
of the filamentation mode in our simulations is an upper limit to the growth rate of any parallel mode that 
we do not see, so {in fact} all our simulations have an observed growth rate
$\Im\omega_{obs}\lesssim 0.3\, \Omega_i$ and some simulations show
$\Im\omega_{obs}< 0.2\, \Omega_i$. 
 
If $\Im\omega_{}\ll \Omega_i$ must be strictly maintained, then what are the constraints on the parameters
so the parallel plane-wave mode can play a role? Assuming efficient Bohm-type diffusion, the upstream plasma 
is swept up by the shock on a timescale $c\,r_{CRg}/3\,v_{sh}^2$, so efficient growth 
requires the growth time be substantially smaller than this:
\begin{equation}
1\gg {{\Im\omega}\over {\Omega_i}} \gg \ {{3\,\beta_{sh}^2}\over \sqrt{\gamma_{CR}^2 -1}}.
\label{limit2}
\end{equation}
If conditions allow a rapid growth of the instability on a timescale much shorter than
the shock-capture timescale, {\mkpa meaning if the instability is astrophysically relevant},
then the instability will also evolve in an 
environment that is essentially not changed by the inflow of fresh material, implying that our
simulation setup using periodic boundary conditions on all sides is appropriate.

For young SNRs the forward-shock speed is $\beta_{sh}\gtrsim 0.01$, so that only a narrow range of 
interesting parameters exists, unless we consider very high-energy cosmic rays for which the 
instability-driving
current is small. We can use Eq.~\ref{limit1} to rewrite the second relation in Eq.~\ref{limit2} 
using $U_{CR}\simeq N_{CR}\,\sqrt{\gamma_{CR}^2 -1}\,m_i\,c^2$ as the cosmic-ray energy density, and
$U_{bulk}={1\over 2}\,N_i\,m_i\,v_{sh}^2$ as the bulk kinetic energy density of the upstream plasma:
\begin{equation}
U_{CR}\gg 12{{v_A}\over {v_{sh}}}\,U_{bulk}\approx 0.1\,U_{bulk}.
\label{limit3}
\end{equation}
Here the second relation applies to SNRs
with small upstream plasma density $N_i \lesssim 0.1\ {\rm cm^{-3}}$, such as SN~1006, RX~J1713-3946, or 
Vela Junior, for which the Alfv\'en speed is $v_A \simeq 30$~km/s for a reasonable magnetic-field strength 
$B_0 \simeq 5\ \mu$G. It is unlikely that the energy density in cosmic rays is much larger than the bulk
energy density of the upstream plasma, suggesting that this instability
{\mkpa can operate for only a few e-folding times} in those SNRs.
Remnants that expand into a high-density environment may be better suited for a strong growth
of the non-resonant mode, but in those cases one has to consider ion-neutral collisions which can reduce the 
growth of the instability \citep{rev08a}.
 
The relation Eq. \ref{limit3} {\mkpa may be} a more severe constraint
than the requirement 
\begin{equation}
k_{\parallel max}\,r_{CRg} = \frac{\zeta}{2} \frac{v_{sh}^2}{v_{A}^2}
={{U_{CR}}\over {U_{bulk}}}\,{{v_{sh}^3}\over {2\,c\,v_{A}^2}}\gg 1
\end{equation}
for the instability to exist in the first place (compare Eq.~\ref{eq1}), and should be used in
{\mkpa addition to it. Equation 12 also shows that cosmic rays of very high energy are not necessarily 
better triggers of magnetic-field amplification. The relevance of cosmic rays in a certain energy band
primarily depends on the energy density they carry. 

Our simulations used cosmic rays with a moderate 
Lorentz factor $\gamma_{CR}=2$. It may be interesting to speculate how our results might change, had 
we been able to use a substantially higher cosmic-ray Lorentz factor, e.g., $\gamma_{CR}=1000$. A spatial 
re-organization of the cosmic rays was not observed in our simulations, and higher-energy particles are more difficult 
to concentrate in certain locations, so the homogeneity of the cosmic-ray distribution would
most likely not change. Leaving the energy density 
in cosmic rays unchanged, we would expect a much slower growth of turbulence on much larger scales. The 
condition $\Im\omega_{}\ll \Omega_i$ should always be on account of the small number of cosmic rays required to
carry their energy density (see Eq.~\ref{limit1}), and so the parallel mode should be initially excited and then
give way to the filamentation and formation of cavities that was observed in the MHD simulations and our PIC studies.
For the backreaction and saturation we therefore do not expect fundamental deviations from the behavior that we
saw in our simulations. We studied systems with turbulence growth on a variety of scales relative to the plasma scale, 
and we never saw a systematic variation in the saturation mechanism and level, that might indicate a dependence on the
wavelength of the turbulence. Even though we could not possibly study a system with, e.g., TeV-ish cosmic rays, we 
therefore feel
there is no reason to assume that in such a situation the bulk acceleration of the plasma and the cosmic rays
would proceed in a different way. The saturation level of the turbulent magnetic field would then also be similar to what we
find. What remains unclear is the actual equilibrium amplitude in a realistic astrophysical scenario, 
in which the instabilities operate under
a competition of saturation through non-linear backreactions with driving through the influx of fresh material. Since the
saturation level observed in our simulations corresponds to a situation in which the backreactions are relatively fast,
also compared with the initial growth, it appears likely that even under continuous inflow of new plasma the 
equilibrium amplitude of the turbulent magnetic field does not exceed the peak values seen in the PIC simulations, so that
$\delta B/B\sim 1$ remains the most probably result.}

\acknowledgments
Part of the simulations have been performed on Columbia at NASA Advanced Supercomputing (NAS).
Partial support by the National Center for Supercomputing Applications under PHY070013N is acknowledged,
where we used the Tungsten and Mercury systems. The work of J.N. was supported by MNiSW during 2005-2008 
as research 
project 1 P03D 003 29 and The Foundation for Polish Science through the HOMING program, which
is supported by a grant from Iceland, Liechtenstein and Norway through the EEA Financial Mechanism.
K.-I.N. is supported by AST-0506719, HST-AR-10966.01-A, NASA-NNG05GK73G, and 
NASA-NNX07AJ88G.

\appendix
\section{Transformations of the Particle Distributions and their Momenta}\label{spectra}

It is convenient to use cylindrical phase-space coordinates $p_\parallel$, $p_\perp$,
and $\phi$, where the differential volume element is 
$d^3p=dp_\parallel\,dp_\perp\,p_\perp d\phi=\frac{1}{2}dp_\parallel\,dp_\perp^2\,d\phi$,
with the axial component $p_\parallel$ defined, in this case, by the direction of a 
Lorentz transformation. In our discussion the azimuthal angle can be ignored, so it is 
sufficient to consider gyrotropic distribution functions 
\begin{equation}
f(p_\parallel,p_\perp)={1\over \pi}\,{{dn}\over {d^3 x\,dp_\parallel\,dp_\perp^2}}.
\end{equation}
If the distribution function is known in some inertial reference frame $K$, then in
a second reference frame $K'$ moving at velocity $\beta\,c\,{\bf e_\parallel}$ with
respect to $K$, the density, bulk velocity, and total energy density of the particles are given by
\begin{eqnarray}
\label{kpdens} n'&\equiv&\int d^3p'\ f'\left({\bf p'}\right) \\
\label{kpbulk} {\bf V'}&\equiv&\frac{1}{n'}\int d^3p'\  {\bf v'}\, f'\left({\bf p'}\right) \\
\label{kpenergy} w'&\equiv&\int d^3p'\ E'\, f'\left({\bf p'}\right),
\end{eqnarray}
where primed quantities are measured in frame $K'$.
We exploit the invariance of $f$ and $p_\perp$ under the Lorentz transformation to 
rewrite Eq. \ref{kpdens} as
\begin{displaymath}
n'=\pi\,\int\limits_{-\infty}^{\infty} dp'_\parallel \int\limits_0^\infty dp^2_\perp 
\ f\left(p_\parallel\left(p'_\parallel\right),p_\perp\right).
\end{displaymath}
By the appropriate change of coordinates, the effect of the Lorentz transformation can
be confined to the Jacobian of the coordinate transformation:
\begin{equation}\label{kpdens2}
n'=\pi\,\int\limits_{-\infty}^\infty dp_\parallel \ \int\limits_0^\infty dp^2_\perp 
\ f\left(p_\parallel,p_\perp\right)\,\left|\frac{dp'_\parallel}{dp_\parallel}\right|_{p_\perp=\rm const}.
\end{equation}
Now $p'_\parallel=\gamma\,\left(p_\parallel-\beta\,\frac{E\left({\bf p}\right)}{c}\right)$, 
where $E\left({\bf p}\right)\equiv\sqrt{m^2c^4+\left(p^2_\parallel+p^2_\perp\right)c^2}$. Then the Jacobian
is
\begin{equation}\label{jacobian}
\left|\frac{dp'_\parallel}{dp_\parallel}\right|_{p_\perp=\rm const.}
=\gamma\,\left(1-\frac{\beta\,c}{E\left({\bf p}\right)}p_\parallel\right).
\end{equation}
If we {consider} a particle distribution that is isotropic, homogeneous, 
and monoenergetic with scalar momentum $p_0$ and density $n$
in frame $K$, 
\begin{equation}\label{kdist}
f\left({\bf p}\right)=\frac{n}{2\pi p_0}\delta\left(p_\parallel^2+p_\perp^2-p_0^2\right),
\end{equation}
we {will} find that $n'=\gamma\,n$. The Jacobian must also be used when setting up the simulations, because
in their flow frame cosmic rays are supposed to follow a distribution according to Eq.~\ref{kdist}, which 
must be transformed into the simulation frame, which is the upstream rest-frame ahead of the SNR shock.

We can compute the bulk velocity (Eq. \ref{kpbulk}) in a similar manner; we express the
{particle} velocity ${\bf v'}$ in terms of the integration variables, and only the parallel component 
$v'_\parallel$ will survive integration. Now $v'_\parallel
=(v_\parallel-\beta\,c)/(1-\beta\,v_\parallel/c)$, and
$v_\parallel=p_\parallel\,c^2/E$. Then for isotropic and monoenergetic particles (Eq.~\ref{kdist})
we may write Eq.~\ref{kpbulk} as
\begin{equation}\label{kpbulk2}
{\bf V'}={\bf e_\parallel}\frac{\pi\,c }{\gamma\,n} 
\int\limits_{-\infty}^\infty dp_\parallel\ 
\int\limits_0^\infty dp^2_\perp \ 
\frac{p_\parallel\,c -\beta\,E}{E-\beta\,p_\parallel\,c} \,
f\left(p_\parallel,p_\perp\right)\,\left|\frac{dp'_\parallel}{dp_\parallel}\right|_{p_\perp=\rm const}.
\end{equation}

When Eq. \ref{kpbulk2} is solved with the appropriate substitutions (Eqs. \ref{jacobian}
and \ref{kdist}), the bulk velocity is found to be
\begin{equation}\label{bulkvel}
{\bf V'}=-\beta\,c\,{\bf e_\parallel}.
\end{equation}
Thus the velocity for the Lorentz transformation to the rest frame $K$ of the particle distribution
is equal to the bulk, or drift, velocity of the particles in another frame $K'$, such as the
simulation frame. The particle distributions are not necessarily isotropic in any frame of reference,
but Eq.~\ref{kpbulk2} nevertheless allows us to properly calculate the drift velocity of a particle 
population and to transform the particle distribution into the flow rest-frame, so that the
anisotropy properties can be investigated. We can also distinguish bulk and 
random momentum, and the transformed distribution function in the flow rest-frame gives the
particle distribution in random momentum that carries information on heating of the particles and any
deviations from the initially prescribed Maxwellians for the background plasma and monoenergetic 
distributions for the cosmic rays.

\begin{figure}
\plottwo{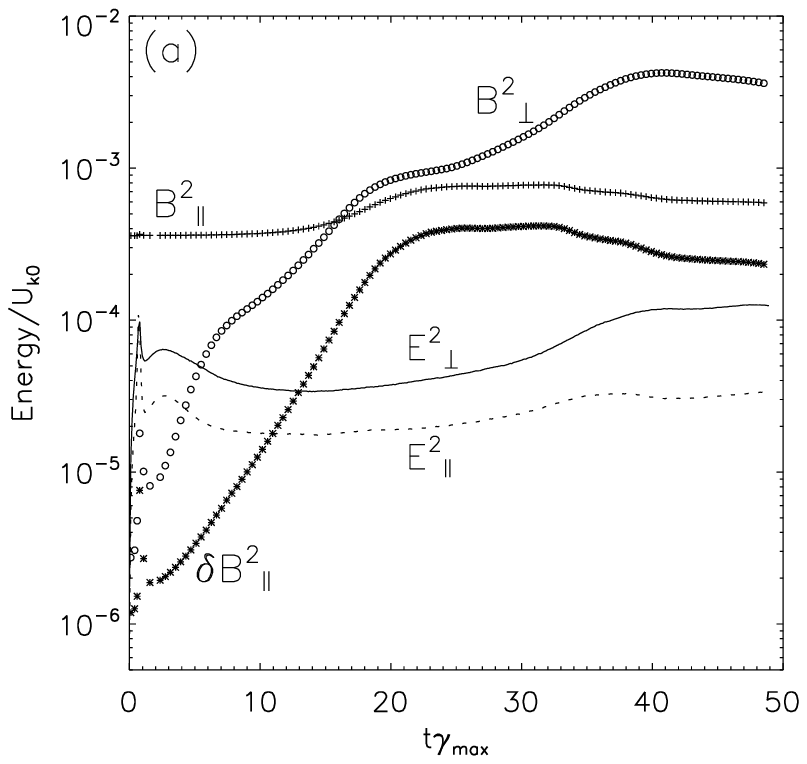}{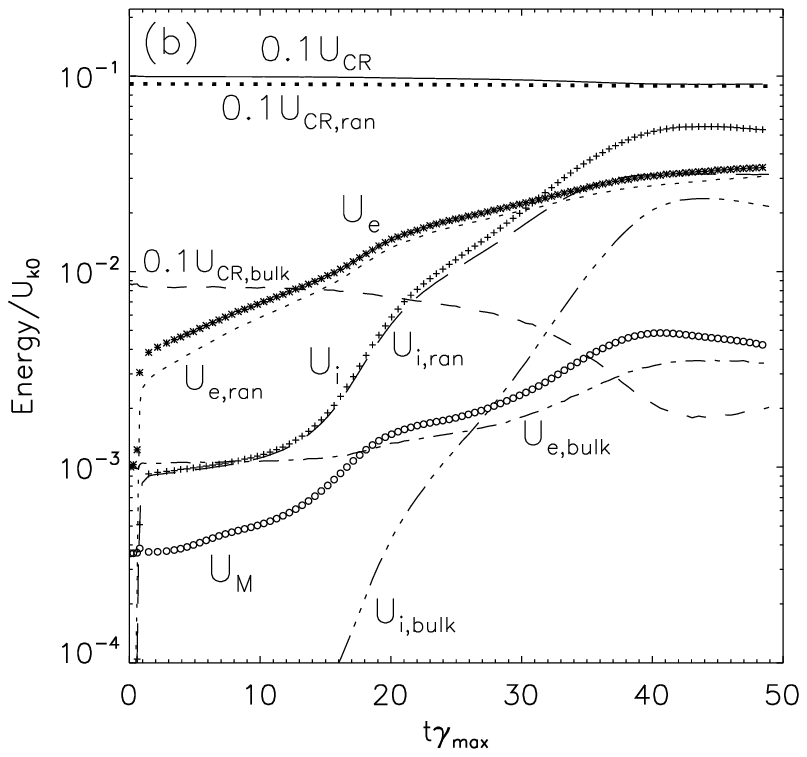}
\caption{{For the main three-dimensional simulation (run A)
using ion-electron mass ratio $m_i/m_e=10$, we show the temporal evolution 
of the average energy density in electromagnetic fields (left panel)
and particles (right panel), both normalized to the 
initial total kinetic energy in the system, $U_{k0}$. 
Time is in units of the {inverse} growth rate $\gamma_{max}^{-1}$ 
for the most unstable mode as predicted in the analytical theory (Eq. 2). 
$B^2_{\perp}$ and  $E^2_{\perp}$ indicate the magnetic and electric field energy
densities in the components perpendicular to the cosmic-ray ion drift direction, {i.e.
$\langle B_y^2+B_z^2\rangle/(2\mu_0)$} and correspondingly for the electric field.
$B^2_{\parallel }$, $\delta B^2_{\parallel}$, and  $E^2_{\parallel}$ are defined analogously,
where $\delta B^2_{\parallel}$ involves only the turbulent component. 
In the right panel, the total energy density of cosmic rays, ambient electrons, and ions 
is split into bulk and random components as outlined in Appendix~\ref{spectra}.
The total kinetic energy density in cosmic-ray ions, $U_{CR}$, 
and its components are scaled by a factor of 10. For comparison, the right panel also shows
the time evolution of the volume-averaged magnetic energy density, 
$U_M=\langle B_x^2+B_y^2+B_z^2\rangle/(2\mu_0)$.} 
 \label{energy}}
\end{figure}

\begin{figure}
\plotone{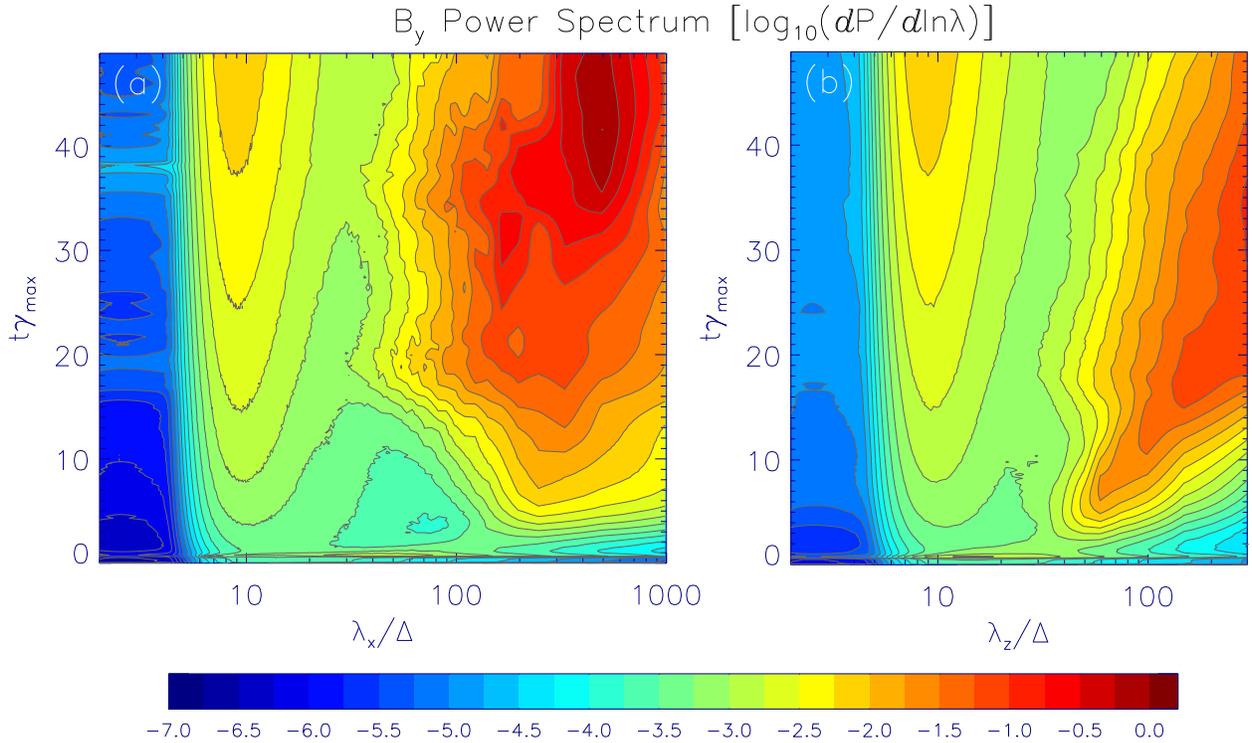}
\caption{The time evolution of the Fourier power spectrum of the {perpendicular 
magnetic-field component $B_y$ in run A}.
The spectra in wavelengths along the drift direction $\lambda_x=2\pi/k_{\parallel}$  {are shown
in the left panel (a), whereas spectra for the direction perpendicular to both
the drift velocity {and $B_y$}
are displayed in the right panel (b).} Note that at small wavelengths ($\lambda\le 5\Delta$) 
{strong} filtering reduces the Fourier power to very small values. 
The spectra are normalized to the peak spectral density.
\label{fBy}}
\end{figure}

\begin{figure}
\plotone{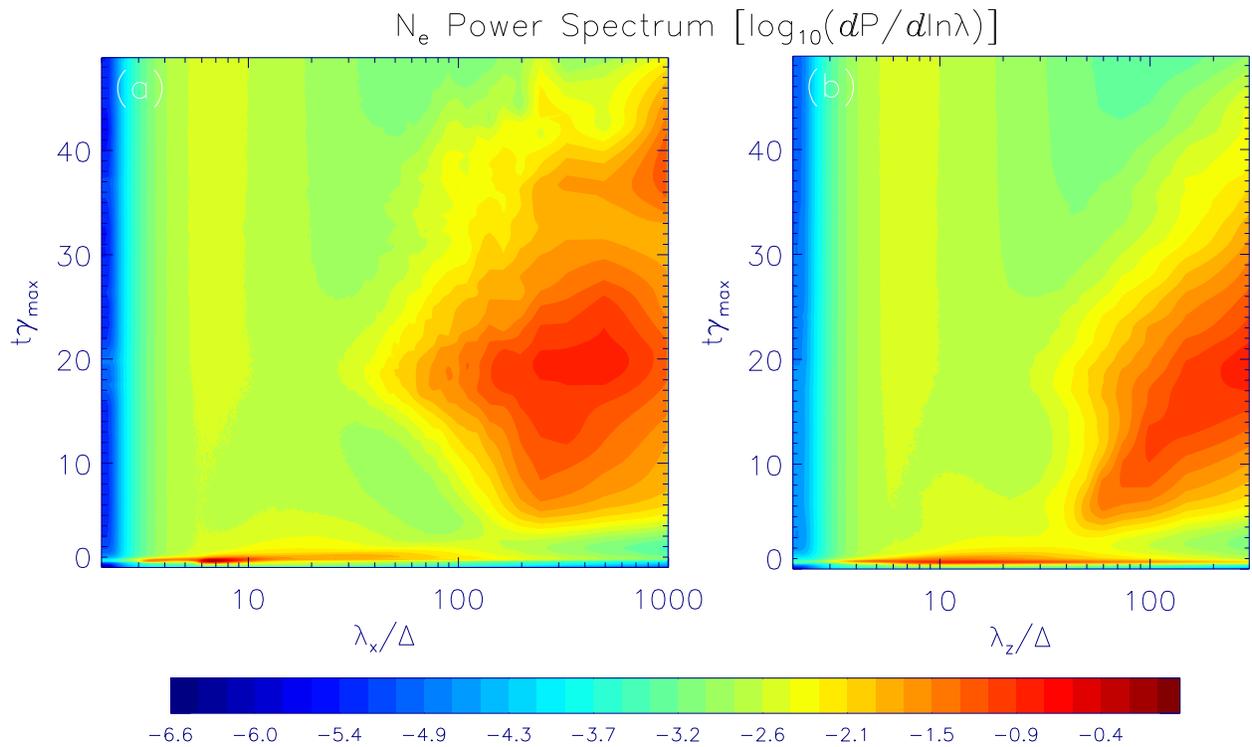}
\caption{The temporal evolution of the electron-density power spectrum for run A. 
The spectra are {set up as described in Fig.~\ref{fBy} and} normalized to the peak spectral density.
\label{fNe}}
\end{figure}

\begin{figure}
\plotone{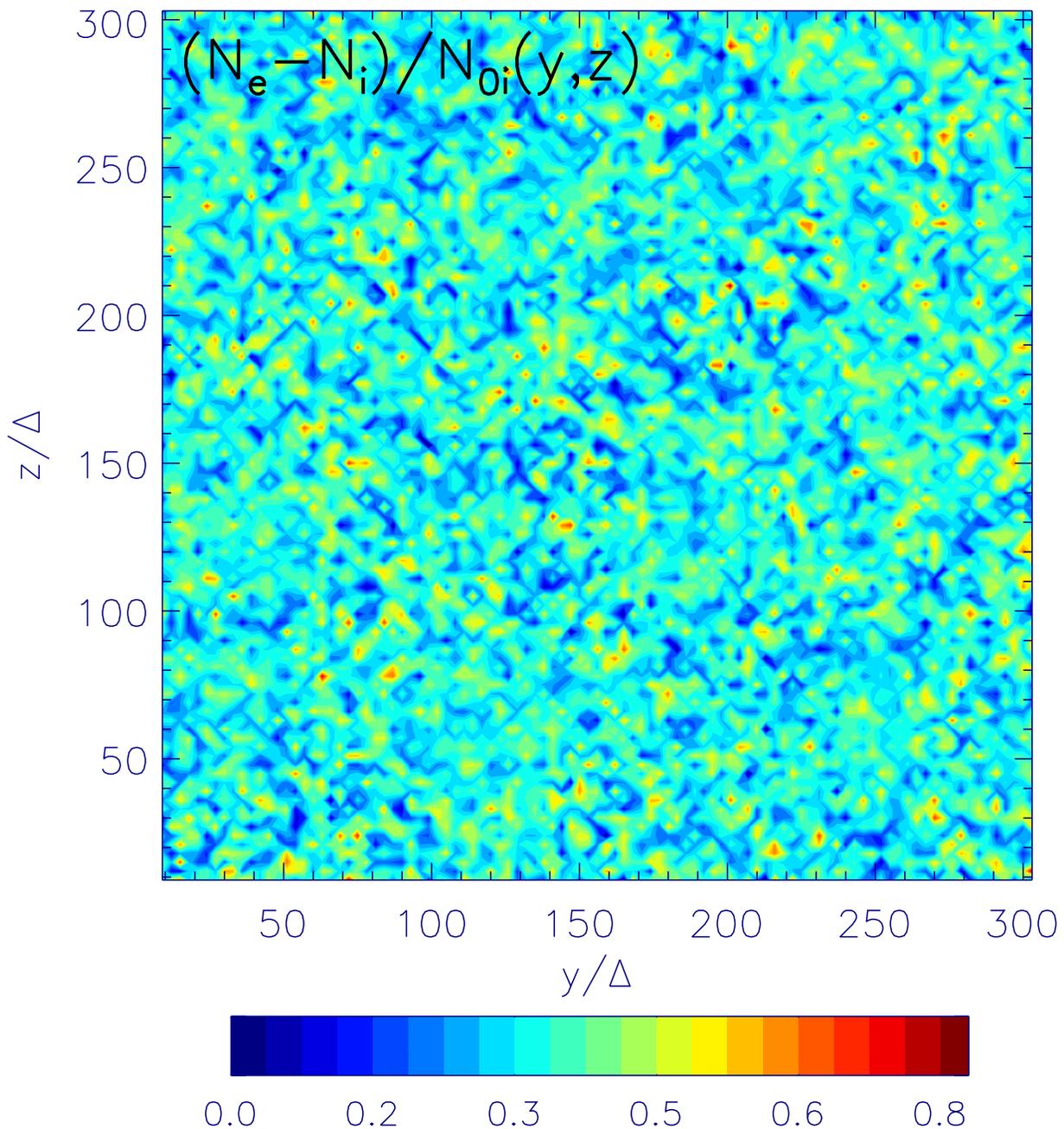}
\caption{Ambient electron-ion density contrast $N_e-N_i$ normalized to the initial
{ion density for {run A at $x/\Delta=500$ and $t\simeq 10\gamma_{max}^{-1}$, as in Fig. 
\ref{nib10}}. The average value for the normalized density contrast is $1/3$ on
account of the excess electrons that are needed to neutralize the charge carried by the
cosmic-ray ions. There are no systematic large-scale deviations from the mean, indicating
that the electron} and ion density distributions are nearly cospatial. 
\label{ne-ni10}}
\end{figure}

\begin{figure}
\plottwo{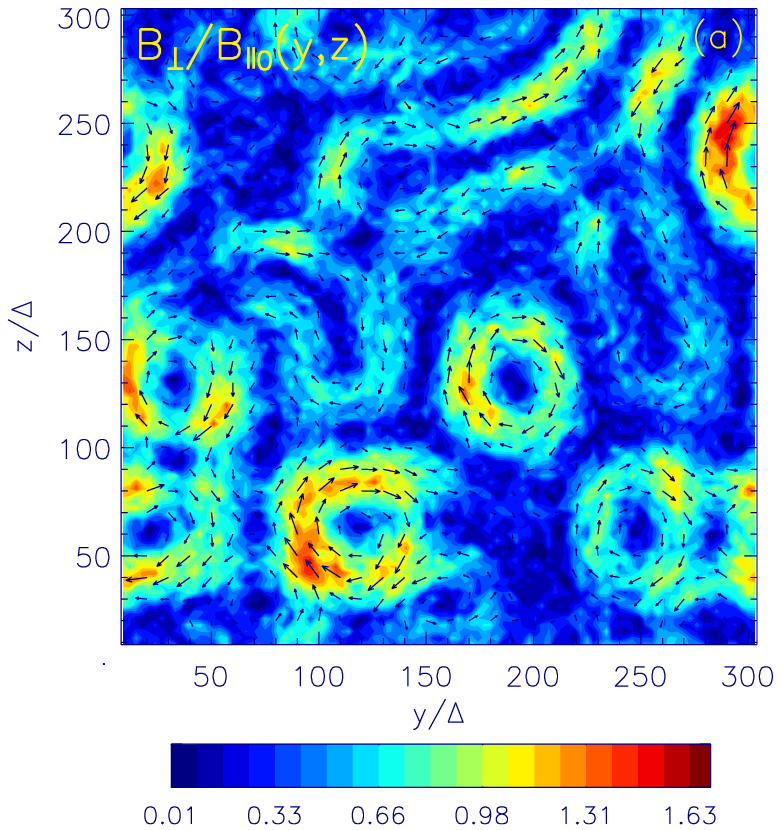}{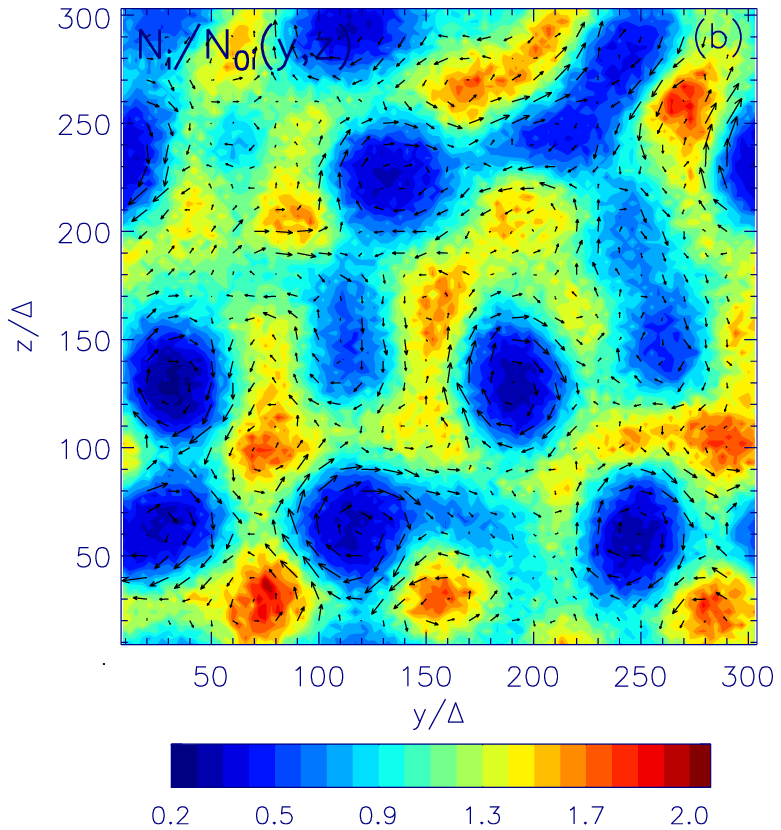}
\caption{{The left panel displays the magnitude and direction, indicated by the arrows,} 
of the perpendicular magnetic field component $B_{\perp}=(B^2_y+B^2_z)^{1/2}$ in the
 plane perpendicular to the cosmic-ray drift direction at the {grid} position $x/\Delta=500$
 and time $t\simeq 10\gamma_{max}^{-1}$. $B_{\perp}$ is normalized to the amplitude
 of the homogeneous field $B_{\parallel 0}$. The right panel shows the density of ambient ions,
$N_i$, in units of the initial ion density at the same location and time.
The electron distribution follows that of ambient ions (see Fig. \ref{ne-ni10}).
\label{nib10}}
\end{figure}

\begin{figure}
\plotone{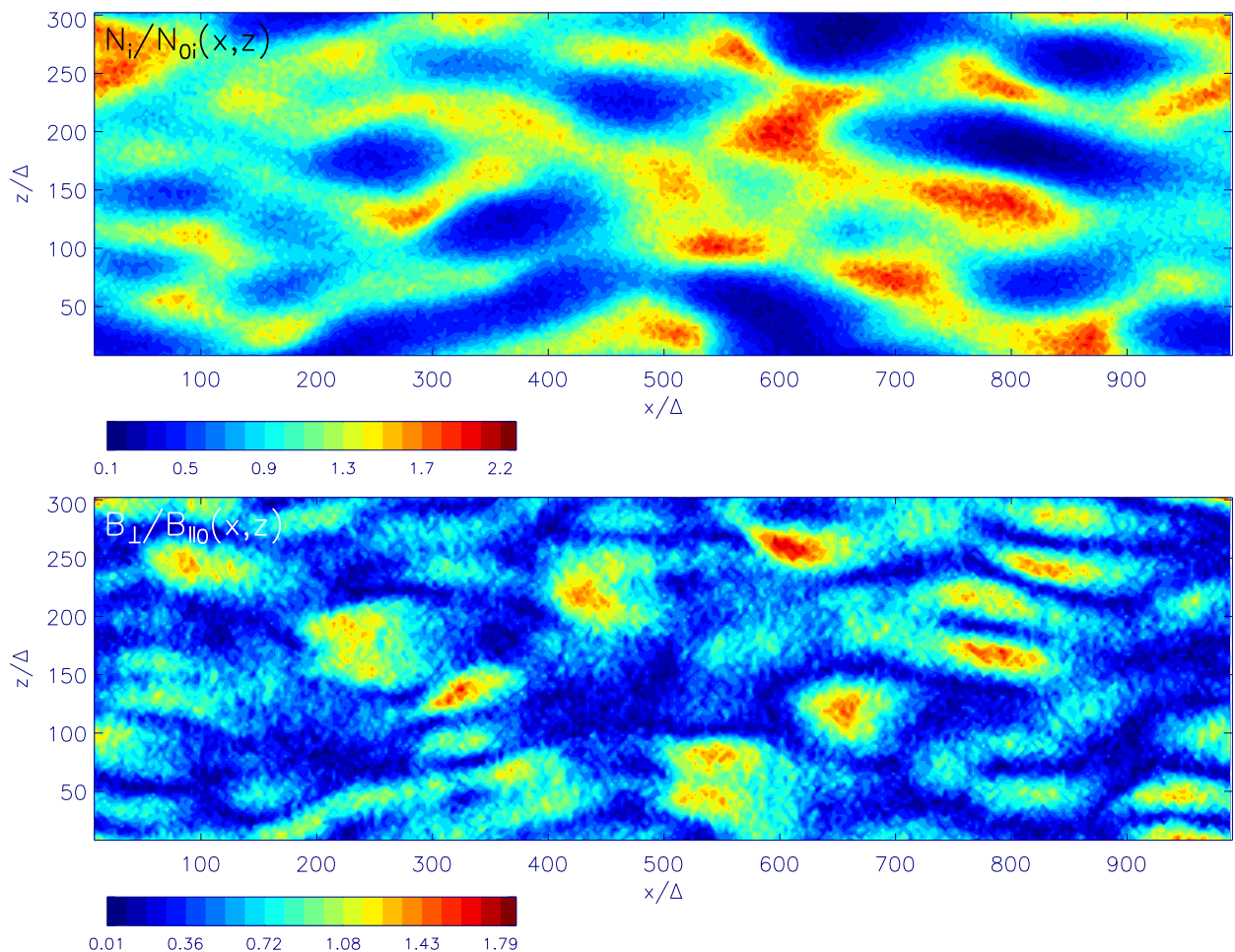}
\caption{The magnitude and direction of the perpendicular magnetic field component 
 $B_{\perp}=(B^2_y+B^2_z)^{1/2}$ (bottom panel) and the ambient ion density $N_i$ (top panel) in the
 plane of the cosmic-ray ion drift direction at $y/\Delta=150$
 and $t\approx 10\gamma_{max}^{-1}$. {A comparison with Fig.~\ref{nib10} illustrates the appearance 
of oblique filamentary structures.} 
\label{nibXZ10}}
\end{figure}

\begin{figure}
\plotone{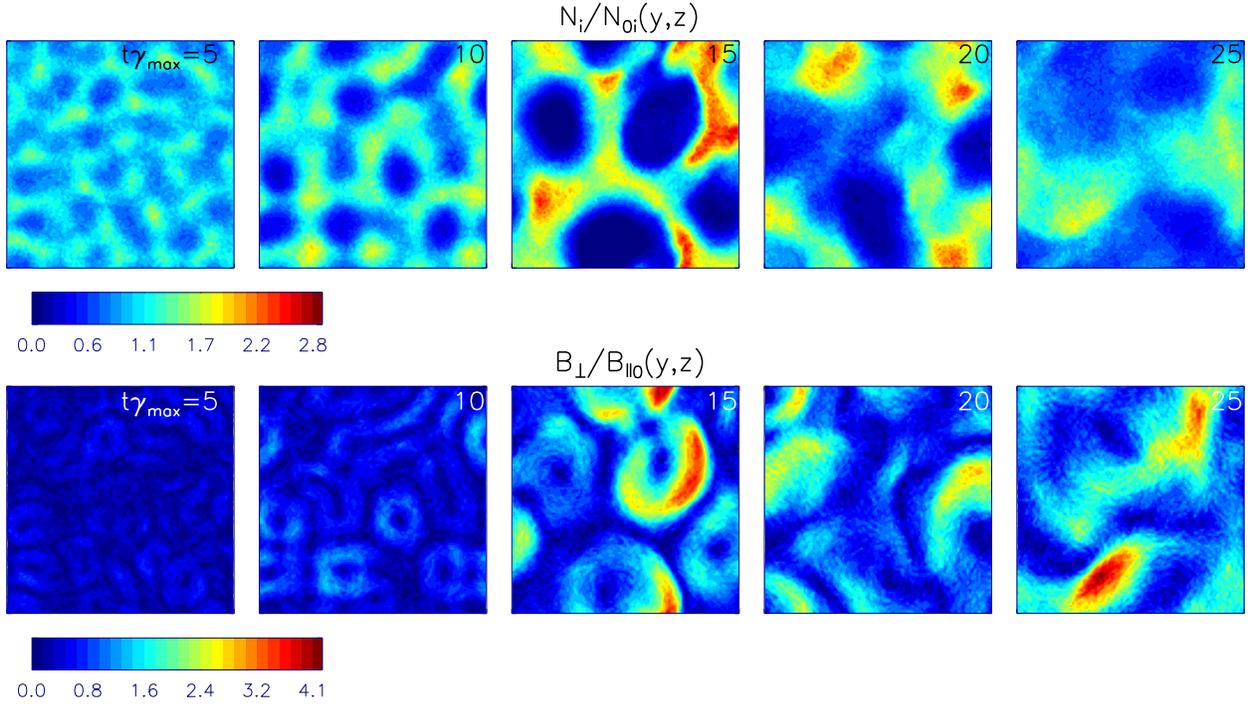}
\caption{Snapshots of the time evolution of the ion density and the perpendicular magnetic-field component
structures in the plane perpendicular to the cosmic-ray ion drift direction at $x/\Delta=500$.
\label{multi}}
\end{figure}

\begin{figure}
\plottwo{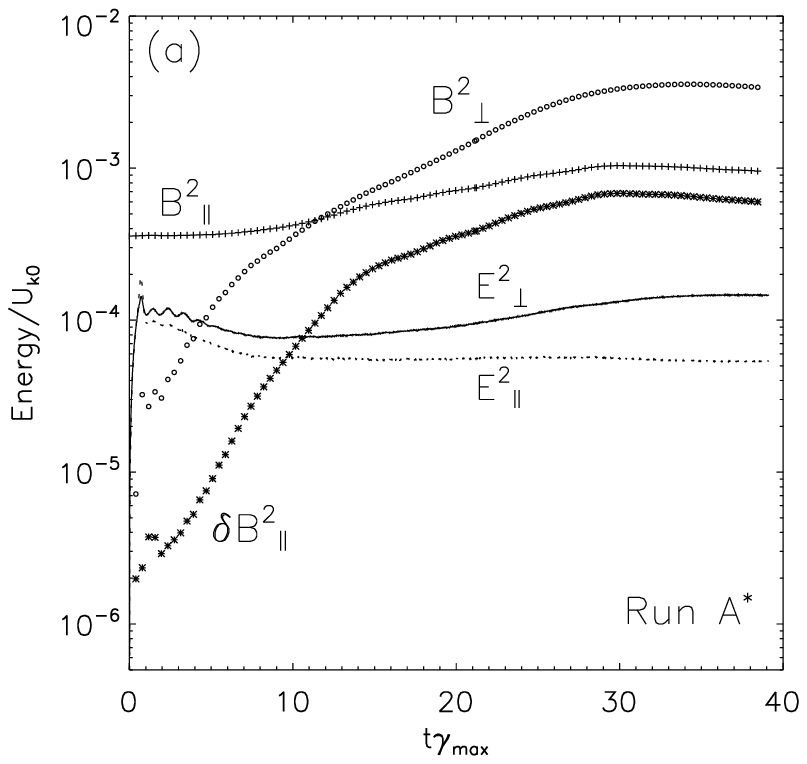}{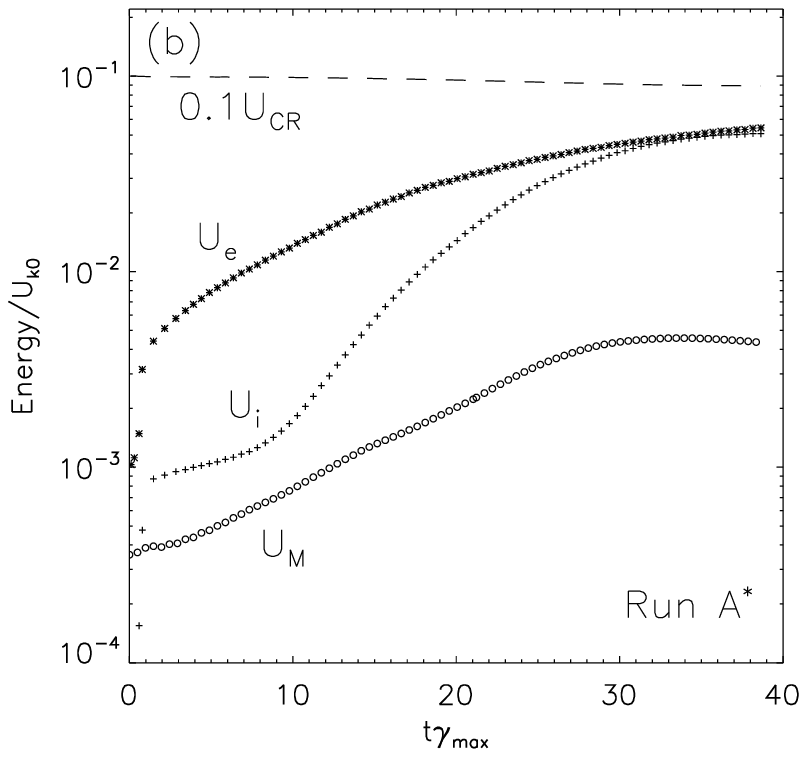}
\caption{Temporal evolution of the total energy density in electromagnetic fields and particles 
for run A$^\ast$ (see Fig. \ref{energy}).
 \label{energy2-D}}
\end{figure}

\begin{figure}
\plotone{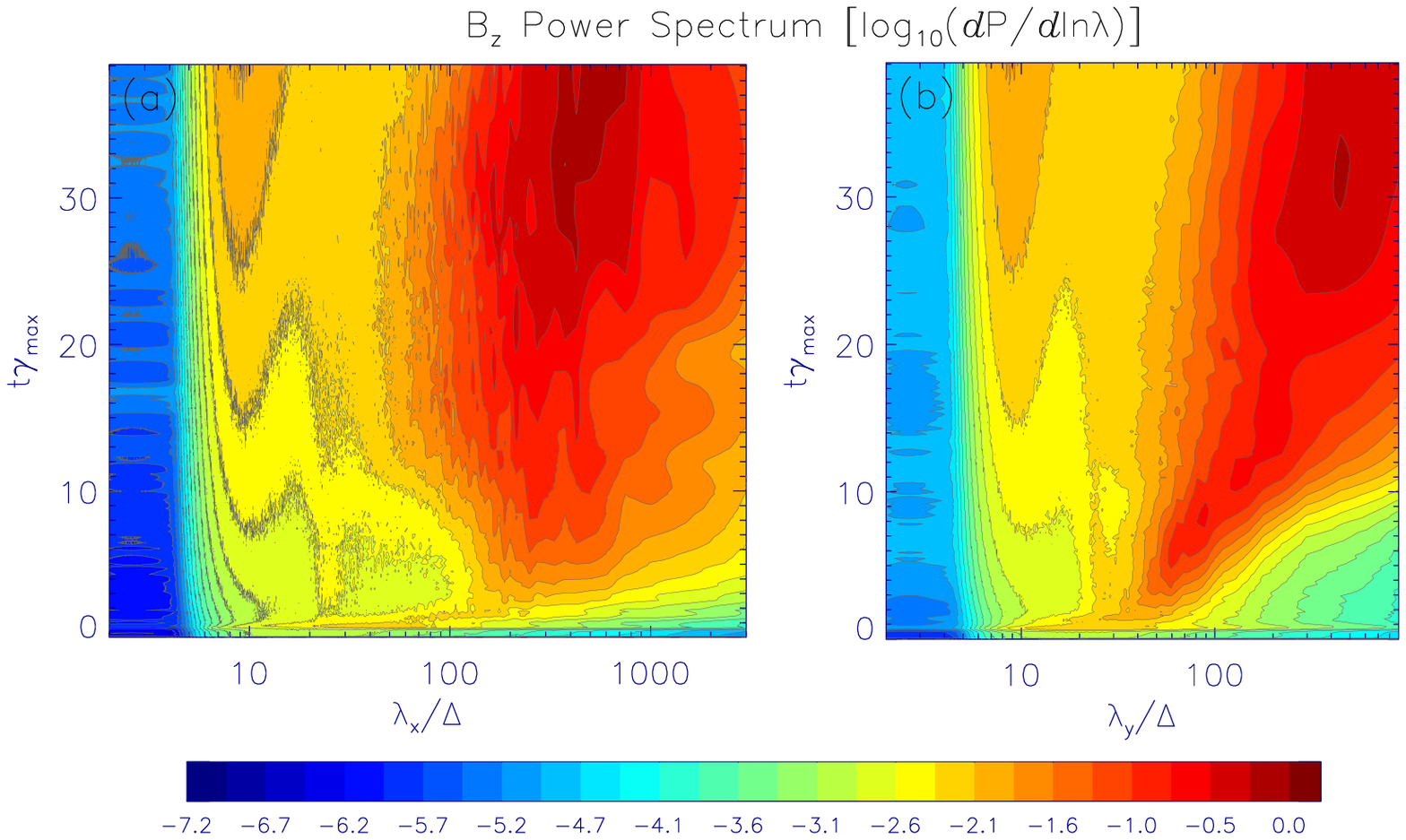}
\caption{Temporal evolution of the {Fourier} power spectrum of the magnetic field component 
$B_z$ for run A$^\ast$ (see Fig. \ref{fBy}).
\label{fBy2-D}}
\end{figure}

\begin{figure}
\plottwo{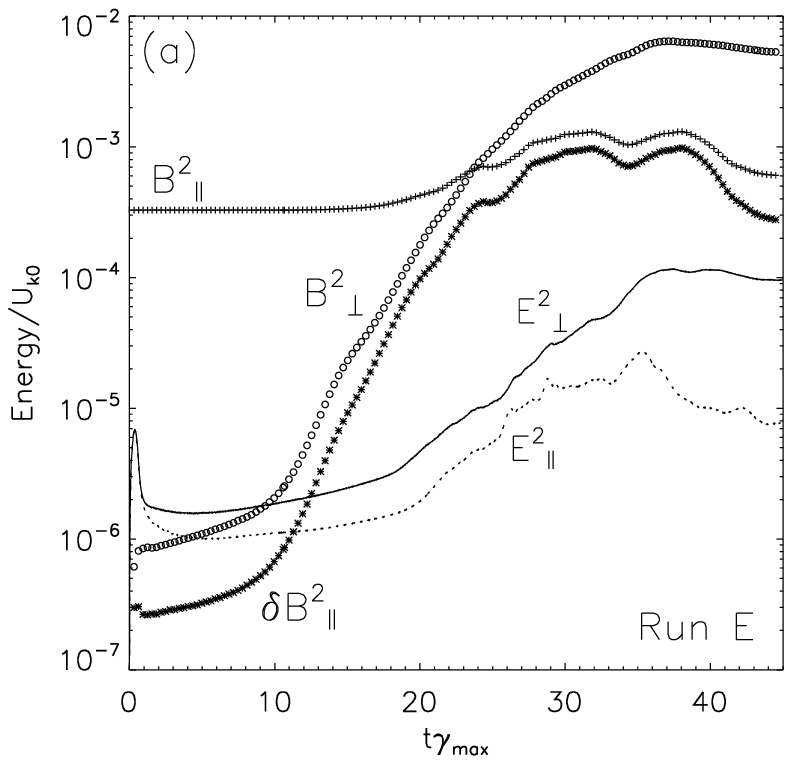}{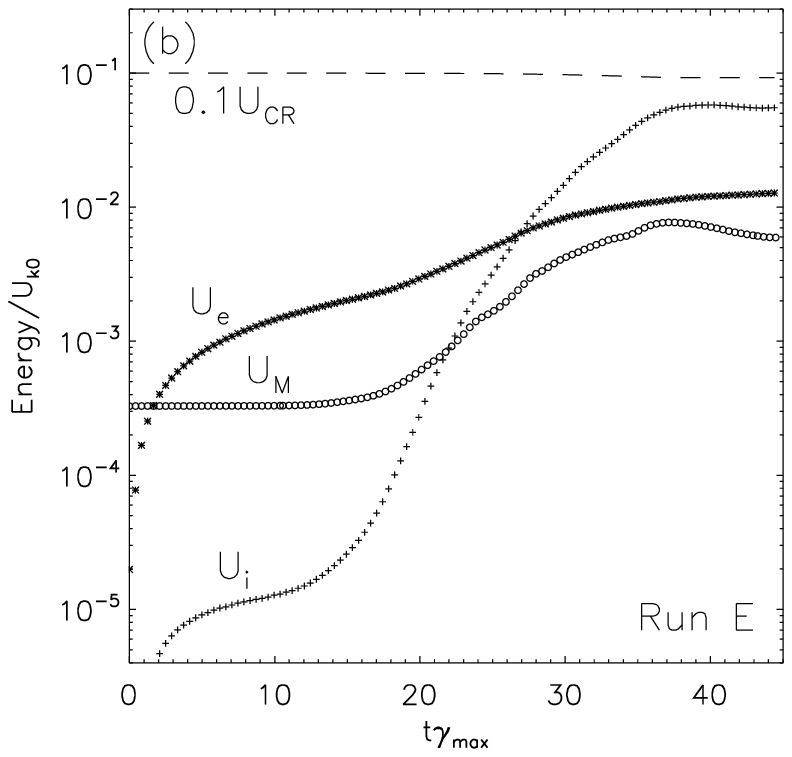}
\caption{Temporal evolution of the total energy density in electromagnetic fields and particles for the 
two-dimensional simulation with a realistic mass ratio $m_i/m_e=500$ (run E; see Fig. \ref{energy}).
 \label{energy500}}
\end{figure}

\begin{figure}
\plottwo{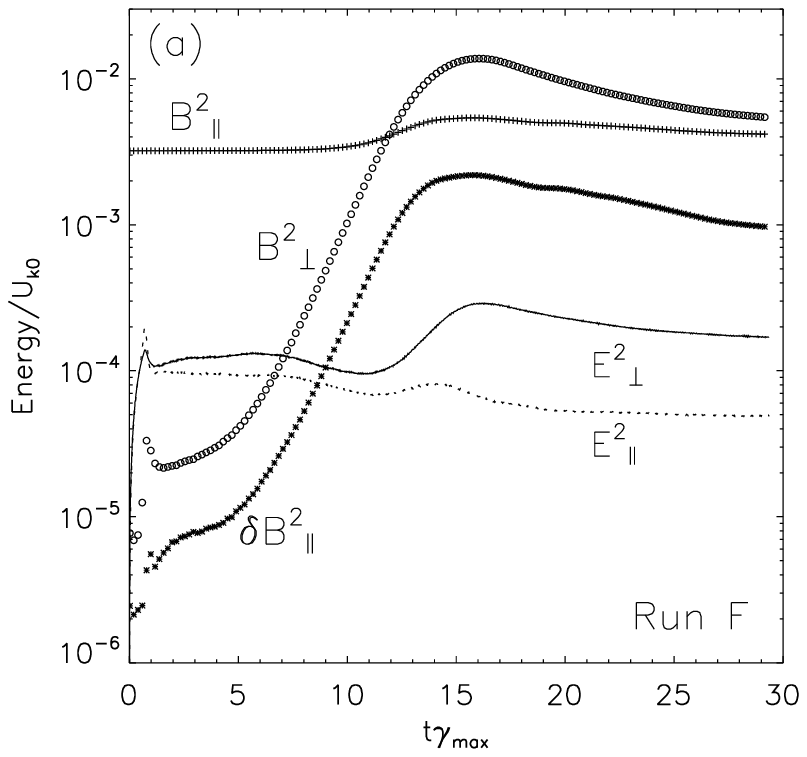}{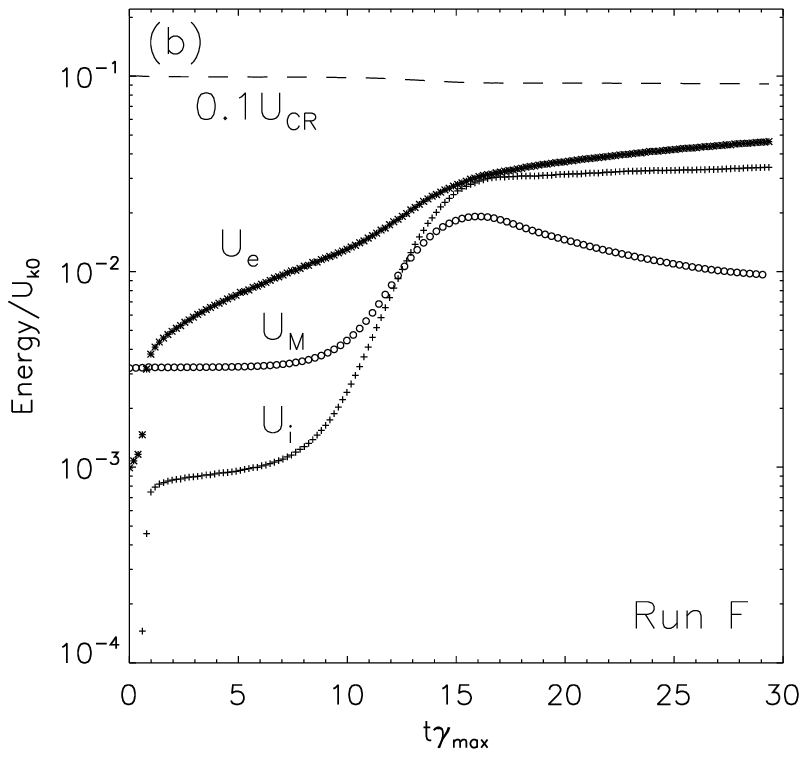}
\caption{Temporal evolution of the total energy density in electromagnetic fields and particles for 
the two-dimensional simulation assuming $\lambda_{max}=150$ and $m_i/m_e=10$ (run F; see Fig. \ref{energy}).
 \label{energy150}}
\end{figure}

\begin{figure}
\plotone{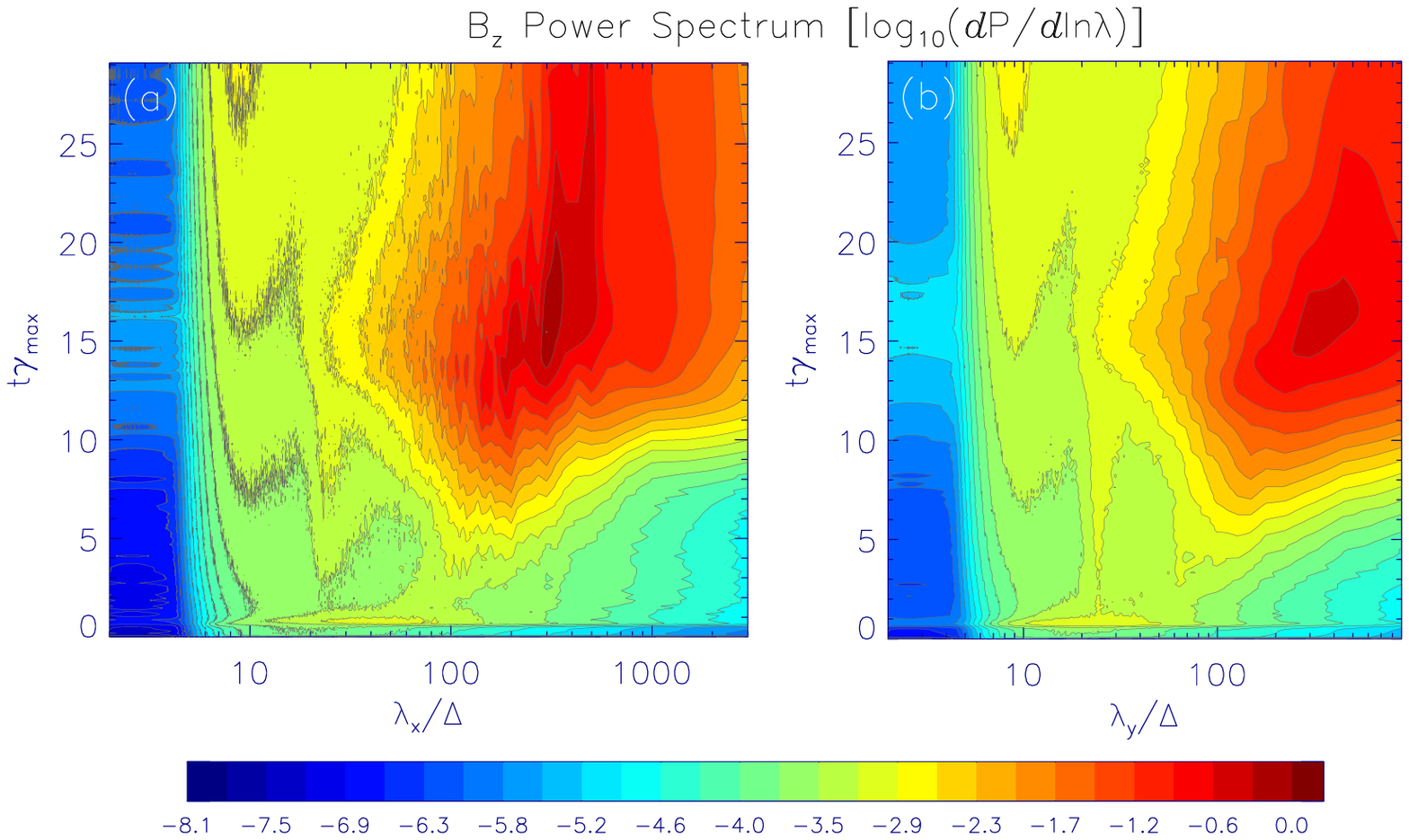}
\caption{Temporal evolution of the {Fourier} power spectrum of the magnetic field component 
$B_z$ for run F (see Fig. \ref{fBy}).
\label{fBy150}}
\end{figure}

\begin{figure}
\plotone{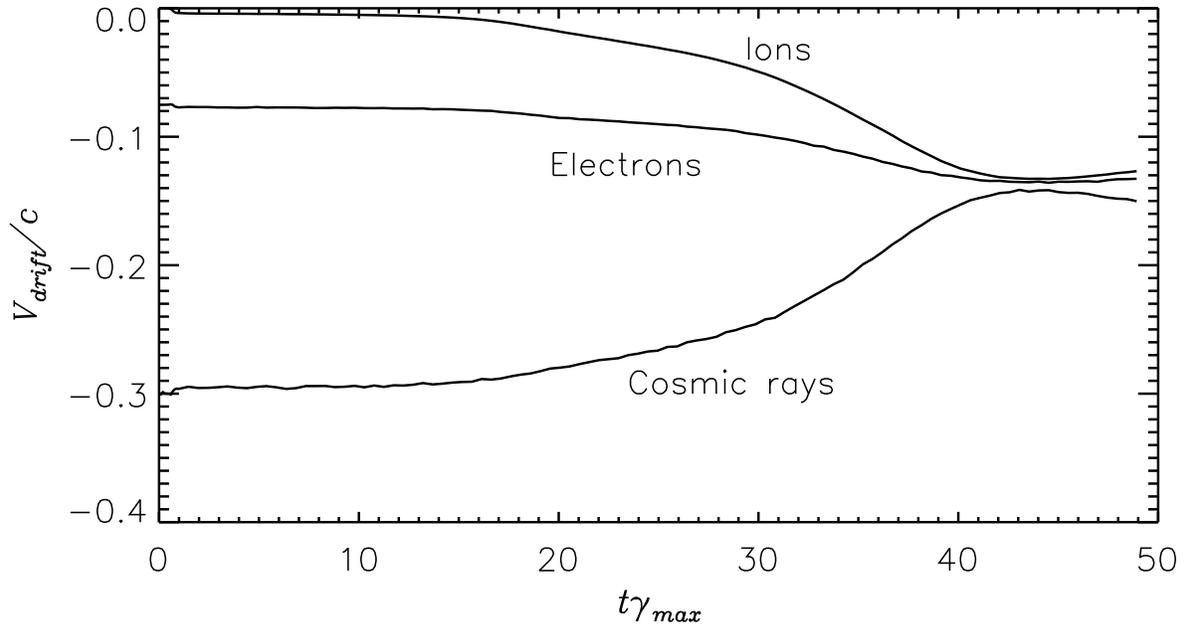}
\caption{The time evolution of the bulk (average) velocity of all particle species for run A.
\label{f-drift}}
\end{figure}

The total energy density $w'$ is easy to calculate as well, but it is useful to
separate it into rest-mass, random and bulk kinetic energy densities as {
$w'\equiv n'\,m\,c^2+U'_{\rm ran}+U'_{\rm bulk}$. For
simplicity we define them as $U'_{\rm ran}\equiv (w'/\gamma)-n'\,m\,c^2$ and 
$U'_{\rm bulk}\equiv (\gamma-1)\,w'/\gamma$, as
a more detailed calculation would yield for an isotropic distribution. Fig.~\ref{energy}
shows the random and bulk kinetic energy densities (without primes) for the main three-dimensional
experiment, run A. Note that the drift velocities, and hence $\gamma$, continuously 
change throughout the simulation.}

\clearpage

\begin{deluxetable}{lcccccccccccc}
\tablecaption{Simulation Parameters and Results}
\tablewidth{0pt}
\rotate
\tablehead{
Run & Grid & $t^{max}$ & $m_i/m_e$ & $N_i/N_{CR}$ & $\lambda_{si}$ & $\lambda_{max}$ 
    & $v_{sh}/v_{A}$ & $\omega_{pe}/\Omega_e$ & $\gamma_{max}/\omega_{pe}$ & $\gamma/\gamma_{max}$ 
    & $\theta$ & $\delta B_{\perp}^{max} / B_{\parallel 0}$\\
    & ($\Delta^3$) & ($\gamma_{max}^{-1}$) & & & ($\Delta$) & ($\Delta$) & & & & & ($^{\circ}$) &}
\startdata
A          & 992$\times$304$\times$304 & 48.9 & 10 & 3 & 25.3 &  50 & 19.1 & 22.1 & 0.0137 & 0.2 & 79 & 3.43\\
A$^{\ast}$ & 3000$\times$900$\times$3  & 39.1 & 10 & 3 & 25.3 &  50 & 19.1 & 22.1 & 0.0137 & 0.2 & 79 & 3.16\\
B          & 992$\times$400$\times$400 & 10.5 & 40 & 3 & 50.7 & 100 & 19.3 & 11.6 & 0.0069 & 0.12 & 62 & -\\
C          & 992$\times$400$\times$400 & 16.3 & 150 & 3& 98.2 & 200 & 18.7 & 5.8 & 0.0035 & 0.06 & 68 & -\\
D          & 3000$\times$900$\times$3  & 49.5 & 100 & 3& 80.8 & 150 & 20.3 & 7.8 & 0.0043 & 0.2 & 73 & 4.59\\
E          & 3000$\times$1500$\times$3 & 44.9 & 500 & 3 & 180.8 & 360 & 18.9 & 3.3 & 0.0019 & 0.23 & 68 & 4.43\\
F          & 3000$\times$900$\times$3  & 29.4 & 10  & 3 & 25.3 & 150 & 6.3  & 7.4 & 0.0137 & 0.35 & 53 & 2.07\\
G          & 3200$\times$3000$\times$3 & 72.1 & 50  & 10 & 51.9 & 100 & 65.3 & 31.9 & 0.002 & 0.09 & 80 & 8.1\\
H          & 3200$\times$3000$\times$3 & 24.5 & 50  & 10 & 51.9 & 300 & 21.8 & 10.6 & 0.002 & 0.58 & 53 & 4.1\\
I          & 3200$\times$3000$\times$3 & 23.3 & 50  & 10 & 51.9 & 500 & 13.1 & 6.4 & 0.002 & 0.75 & 39 & 3.3\\
I$^{\ast}$ & 2400$\times$2400$\times$3 & 20.8 & 50\tbn{a} & 10 & 51.9 & 500 & 13.1 & 6.4 & 0.002 & 0.75 & 39 & 7.5\\
J          & 3200$\times$3000$\times$3 & 11.7 & 50  & 30 & 51.9 & 500 & 43.5 & 21.3 & 0.0006 & 0.7  & 45 & -
\enddata
\tablecomments{Parameters and selected results of the simulation runs described in this paper. Listed are: 
the system size $(x\times y\times z)$ in units of the grid cell size $\Delta$, 
the run duration $t^{max}$ in units of 
$\gamma_{max}^{-1}$, the ion-electron mass ratio $m_i/m_e$, 
{the density ratio of ambient ions and cosmic rays,}
the ion skindepth $\lambda_{si}$ in units of $\Delta$, 
the wavelength of the {theoretically expected}
most unstable mode $\lambda_{max}$ (in units of $\Delta$, see Eq. 2), the 
Alfv\'enic Mach number $M_A=v_{sh}/v_{A}$ of the {shock}, {the plasma magnetization as given 
by the electron plasma to cyclotron 
frequency ratio,} the maximum growth rate $\gamma_{max}/\omega_{pe}$ 
(see Eq. 2), the {actual measured} growth rate $\gamma$ in units of $\gamma_{max}$, 
the {obliqueness $\theta=\angle(\vec k, \vec v_s)$ in degrees of the actual} dominant turbulence mode,
and the maximum amplitude of the perpendicular 
magnetic-field perturbations $\delta B_{\perp}^{max}$ relative to the homogeneous magnetic field.}
\tablenotetext{a}{Cosmic-ray particles' mass assumed in this run is 2$\times10^8m_i$.}
\end{deluxetable}

\end{document}